# Bright triplet excitons in lead halide perovskites

Michael A. Becker[1,2,*], Roman Vaxenburg[3,*], Georgian Nedelcu[4,5], Peter C. Sercel[6], Andrew Shabaev[3], Michael J. Mehl[7], John G. Michopoulos[8], Samuel G. Lambrakos[8], Noam Bernstein[8], John L. Lyons[8], Thilo Stöferle[1], Rainer F. Mahrt[1], Maksym V. Kovalenko[4,5], David J. Norris[2], Gabriele Rainò[1,4], and Alexander L. Efros[8]

[1]IBM Research – Zurich, Säumerstrasse 4, 8803 Rüschlikon, Switzerland.
[2]Optical Materials Engineering Laboratory, ETH Zurich, 8092 Zurich, Switzerland.
[3]Computational Materials Science Center, George Mason University, Fairfax, VA 22030 USA.
[4]Institute of Inorganic Chemistry, Department of Chemistry and Applied Bioscience, ETH Zurich, 8093 Zurich, Switzerland.
[5]Laboratory of Thin Films and Photovoltaics, Empa – Swiss Federal Laboratories for Materials Science and Technology, CH-8600 Dübendorf, Switzerland.
[6]T. J. Watson Laboratory of Applied Physics, California Institute of Technology, Pasadena, CA 91125 USA.
[7]U.S. Naval Academy, Annapolis, MD 21402 USA.
[8]Center for Computational Materials Science, Naval Research Laboratory, Washington, DC 20375 USA.

*These authors contributed equally to this work.

**Nanostructured semiconductors emit light from electronic states known as excitons[1]. According to Hund's rules[2], the lowest energy exciton in organic materials should be a poorly emitting triplet state. Analogously, the lowest exciton level in all known inorganic semiconductors is believed to be optically inactive. These 'dark' excitons (into which the system can relax) hinder light-emitting devices based on semiconductor nanostructures. While strategies to diminish their influence have been developed[3-5], no materials have been identified in which the lowest exciton is bright. Here we show that the lowest exciton in quasi-cubic lead halide perovskites is optically active. We first use the effective-mass model and group theory to explore this possibility, which can occur when the strong spin–orbit coupling in the perovskite conduction band is combined with the Rashba effect[6-10]. We then apply our model to $CsPbX_3$ (X = Cl, Br, and I) nanocrystals[11], for which we measure size- and composition-dependent fluorescence at the single-nanocrystal level. The bright character of the lowest exciton immediately explains the anomalous photon-emission rates of these materials, which emit ~20 and ~1,000 times faster[12] than any other semiconductor nanocrystal at room[13-16] and cryogenic[17] temperatures, respectively. The bright exciton is further confirmed by detailed analysis of the fine structure in low-temperature fluorescence spectra. For semiconductor nanocrystals[18], which are already used in lighting[19,20], lasers[21,22], and displays[23], these optically active excitons can lead to materials with brighter emission and enhanced absorption. More generally, our results provide criteria for identifying other semiconductors exhibiting bright excitons with potentially broad implications for optoelectronic devices.**

An exciton involves an electron in the conduction band Coulombically bound to a hole in the valence band. Its energy depends in part on the spin configuration of these two charge carriers. In organic semiconductors, the lowest energy exciton is a triplet state in which these two carriers have parallel spins. For the electron and hole to recombine and release a photon, one spin must simultaneously flip to satisfy the Pauli exclusion principle. Because this coordinated process is unlikely, triplet excitons are poorly emitting.

In addition to spin, the exciton energy depends on the atomic orbitals that constitute the conduction and valence bands. In many inorganic semiconductors, the orbital motion and spin of the carriers are strongly coupled. Spin is no longer conserved, and the total angular momentum of each carrier ($J_e$ and $J_h$) must be considered. Further, the exchange interaction mixes these so that only the total exciton momentum $\boldsymbol{J=J_e+J_h}$ is conserved. Due to these and other effects, each exciton state is split into several energy sublevels, known as fine structure. Studies on various materials have found that the lowest energy sublevel is dark. For example, in CdSe, recombination of the lowest exciton requires a change of two units of angular momentum[17]. Because the photon carries one unit, light cannot be emitted unless another unit is simultaneously dissipated, another unlikely process. Thus, despite the added complexity in inorganic semiconductors, they appear to behave like organic semiconductors, *i.e.* exhibiting an optically inactive lowest exciton. Indeed, no exceptions are known, leading to the belief that such states *must* be dark.

We show that this belief is incorrect by examining $CsPbX_3$ (X=Cl, Br, and I) perovskites. Their crystals comprise corner-sharing $PbX_6$-octahedra with $Cs^+$ ions filling the voids between (Fig. 1a). We first approximate the lattice as cubic and calculate band structures (Methods) for $CsPbBr_3$ (Fig. 1b), $CsPbCl_3$, and $CsPbI_3$ (Extended Data Fig. 1). The bandgap occurs at the Brillouin zone's R-point, near which the valence and conduction bands are well described within the effective-mass model (see Supplementary Table 1). The top of the valence band arises from a mixture of Pb 6$s$ and Br 4$p$ atomic orbitals, with an overall $s$ symmetry[24-26]. Thus, including spin, the hole can occupy one of two $s$-like Bloch states with $J_h$=1/2, $i.e.$ $|\uparrow\rangle_h = |S\rangle|\uparrow\rangle$ or $|\downarrow\rangle_h = |S\rangle|\downarrow\rangle$, using standard notation[27,28]. The conduction band consists of Pb 6$p$ orbitals, leading to three possible orthogonal spatial components for the Bloch function: $|X\rangle$, $|Y\rangle$, or $|Z\rangle$ [24-26]. Because of strong spin–orbit coupling, these are mixed with spin to obtain a doubly degenerate $J_e$=1/2 state for the electron at the bottom of the conduction band:

$$|\Uparrow\rangle_e = -\frac{1}{\sqrt{3}}\left[\left(|X\rangle + i|Y\rangle\right)|\downarrow\rangle + |Z\rangle|\uparrow\rangle\right]$$
$$|\Downarrow\rangle_e = \frac{1}{\sqrt{3}}\left[|Z\rangle|\downarrow\rangle - \left(|X\rangle - i|Y\rangle\right)|\uparrow\rangle\right]$$
.  (1)

When the momentum of the electron and hole states are then combined, the exciton splits due to electron–hole exchange into a $J$=0 singlet state,

$$|\Psi_{0,0}\rangle = \frac{1}{\sqrt{2}}\left[|\Downarrow\rangle_e|\uparrow\rangle_h - |\Uparrow\rangle_e|\downarrow\rangle_h\right],  (2)$$

and a threefold degenerate $J$=1 triplet state,

$$|\Psi_{1,-1}\rangle = |\Downarrow\rangle_e|\downarrow\rangle_h, \quad |\Psi_{1,0}\rangle = \frac{1}{\sqrt{2}}\left[|\Downarrow\rangle_e|\uparrow\rangle_h + |\Uparrow\rangle_e|\downarrow\rangle_h\right], \quad |\Psi_{1,+1}\rangle = |\Uparrow\rangle_e|\uparrow\rangle_h,  (3)$$

where each $|\Psi_{J,J_z}\rangle$ is labeled with $J_z$, the $z$-projection of $J$. The probability of light emission due to electron–hole recombination from these excitons can then be calculated (Supplementary Section 1). We find a probability of zero for $|\Psi_{0,0}\rangle$ and nonzero for $|\Psi_{1,J_z=0,\pm1}\rangle$, indicating a dark singlet and bright triplet.

These selection rules are confirmed by group theory. At the R-point, the band-edge electron and hole states transform as irreducible representations $R_6^-$ and $R_6^+$, respectively (superscript denoting parity)[28,29]. Exchange then splits the exciton into a dark singlet ($R_1^-$) and a bright triplet ($R_4^-$). (See Supplementary Section 2 and Supplementary Table 3.)

Detailed calculations (Supplementary Section 1) can then reveal the energetic order of these levels. If only short-range exchange is included, the singlet lies below the triplet (Fig. 1c). However, $CsPbX_3$ perovskites should also exhibit a large Rashba effect[6]. This occurs in semiconductors with strong spin–orbit coupling and an inversion asymmetry. For the closely related hybrid organic–inorganic perovskites, the impact of this effect on photovoltaic and spintronic devices has been heavily discussed[6-10]. Although the cause of the inversion asymmetry (cation positional instabilities[30] or surface effects[10]) remains unknown, the Rashba effect should alter the fine structure. Indeed, the bright triplet exciton can be lowered below the dark singlet exciton.

To examine this possibility, we studied colloidal nanocrystals of $CsPbX_3$ (Methods). Compared to bulk crystals, nanocrystals allow the additional effect of system size to be investigated. Such particles are roughly cube-shaped with edge lengths $L$=8-15 nm (Fig. 1d). Before these were introduced[11], all technologically relevant semiconductor nanocrystals exhibited slow sub-microsecond radiative lifetimes at cryogenic temperatures due to the lowest exciton being dark[17]. In contrast, $CsPbX_3$ nanocrystals emit ~1000x faster (with sub-nanosecond lifetimes)[12]. Figure 2a shows photoluminescence decays for individual $CsPbI_3$, $CsPbBr_3$, and $CsPbBr_2Cl$ nanocrystals. (All spectroscopy herein was performed at 5 K.) The decay times are 0.85, 0.38, and 0.18 ns, respectively, decreasing with increasing emission energy. Figure 2b presents a larger set of decay times (squares) for individual $CsPbI_3$, $CsPbBr_3$, and $CsPbBr_2Cl$ nanocrystals. All are much shorter than those reported for CdSe,

CdS, CdTe, InAs, InSb, InP, PbSe, PbS, and PbTe nanocrystals[13-16], consistent with the lowest exciton being the bright triplet.

However, fast decays could also indicate emission from trions (charged excitons). Trions are optically active and suffer from nonradiative Auger recombination. In our single-nanocrystal experiments, any trion contribution is already reduced by spectral filtering (Extended Data Fig. 2). To completely eliminate the trion contribution, the photon stream from individual nanocrystals can be analyzed[31]. By correlating emission intensity with lifetime (Fig. 2c,d), pure exciton contributions can be extracted. We confirm fast exciton lifetimes (1.2 and 0.4 ns, respectively) for $CsPbI_3$ and $CsPbBr_3$ nanocrystals, values consistent with ensemble measurements (Extended Data Fig. 3).

To compare with theory, we calculated radiative lifetimes for perovskite nanocrystals within the effective-mass model. In addition to the wavefunctions in equations (2)-(3), exciton confinement within the nanocrystal must be included via envelope functions for the electron and hole. If $CsPbX_3$ nanocrystals were spherical, excitonic lifetimes could be calculated with prior methods (Supplementary Section 3). However, for cubes, the electric field of a photon not only changes across the nanocrystal boundary due to dielectric screening (as in spherical nanocrystals), but it also becomes inhomogeneous[32] (Fig. 2e). We included this inhomogeneity, along with the Rashba effect, the orthorhombic lattice distortion in $CsPbX_3$ nanocrystals[33], and the possibility that their shape is slightly elongated (which leads to tetragonal or orthorhombic symmetry). Only when the Rashba effect was included could a self-consistent model for $CsPbX_3$ nanocrystals be obtained, as now described.

The Rashba coefficient was estimated from low-temperature photoluminescence spectra (see below). If the effective Rashba field is parallel to one of the orthorhombic symmetry axes of the nanocrystal (see Supplementary Section 1 for details and other cases), the bright triplet exciton ($J$=1) is split into three nondegenerate sublevels:

$$\left|\Psi_x\right\rangle = \frac{1}{\sqrt{2}}\left[\left|\Uparrow\right\rangle_e\left|\uparrow\right\rangle_h - \left|\Downarrow\right\rangle_e\left|\downarrow\right\rangle_h\right], \quad \left|\Psi_z\right\rangle = \left|\Psi_{1,0}\right\rangle, \quad \left|\Psi_y\right\rangle = \frac{1}{\sqrt{2}}\left[\left|\Uparrow\right\rangle_e\left|\uparrow\right\rangle_h + \left|\Downarrow\right\rangle_e\left|\downarrow\right\rangle_h\right], \quad (4)$$

which lie below the dark singlet (Fig. 1c). The triplet states represent three linear dipoles polarized along the orthorhombic symmetry axes ($x$, $y$, $z$). Transitions from these three sublevels have the same oscillator strength. Moreover, in cube-shaped nanocrystals, these states still emit as linear dipoles despite the inhomogeneous field (Supplementary Sections 1 and 3).

The triplet exciton radiative lifetime, $\tau_{ex}$, can then be evaluated from[34]:

$$\frac{1}{\tau_{ex}} = \frac{4\omega n E_p}{9 \cdot 137 m_0 c^2} I_{\parallel}^2. \quad (5)$$

with the angular transition frequency, $\omega$, the refractive index of the surrounding medium, $n$, the free-electron mass, $m_0$, the speed of light, $c$, the Kane energy, $E_p = 2P^2/m_0$, and the Kane parameter, $P$ (ref. 27). $I_{\parallel}$ is an overlap integral that includes the electron and hole envelope functions and the field-averaged transition-dipole moment (Supplementary Section 3).

Figure 2b presents the calculated $\tau_{ex}$ for $CsPbX_3$ nanocrystals (circles). The results can be divided into three regimes, depending on the nanocrystal size[35]. When the nanocrystal is smaller than the exciton Bohr radius $a_B$ (strong confinement, orange circles), the predicted radiative lifetime decreases from 2 to 1 ns with increasing emission energy. For large nanocrystals in the opposite limit (green circles), the lifetime should be even shorter as weakly confined excitons exhibit larger oscillator strengths[36]. In this size regime ($L$~15-25 nm), the calculated lifetimes decrease below 100 ps for $CsPbBr_3$ and $CsPbCl_3$ nanocubes. The lifetime would be decreased further in spheres of the same volume (lower inset, Fig. 2e).

The measured photoluminescence decays in Fig. 2b (squares) lie between those predicted for strong and weak confinement. Because the nanocrystal size and $a_B$ are comparable, the electron and hole motion is correlated. If this effect is added (intermediate confinement, blue circles), calculations for $L$~4-16 nm (Supplementary Section 3) agree well with experiment.

The above calculations depend on knowing the Rashba coefficient $\alpha_R$. This was estimated from photoluminescence spectra of individual nanocrystals, which reveal the fine structure directly. Our nanocrystals exhibit one, two, or three peaks, all with near-linear polarization (Fig. 3a-c and Extended Data Figs. 4 and 5). This is consistent with the three nondegenerate exciton sublevels in equation (4) under orthorhombic symmetry, which should emit as orthogonal linear dipoles. The Rashba coefficient ($\alpha_R = 0.38$ eV Å) required to fit the observed splittings (~1 meV) is reasonable, lying between values for conventional III-V quantum wells and organic–inorganic perovskites (see Supplementary Section 1.F). We note that for nanocrystals with tetragonal symmetry, $\left| \Psi_x \right\rangle$ and $\left| \Psi_y \right\rangle$ in equation (4) remain degenerate (Supplementary Section 1.E), explaining another recent observation[37].

Emitting dipoles that are perpendicular (parallel) to the observation direction should show strong (no) emission. Thus, the intensity from each bright-triplet sublevel is explained by both its thermal population and the nanocrystal orientation. Single-line spectra (Fig. 3a) arise when the two upper sublevels are unpopulated. Strong linear polarization from this single line (Fig. 3a, inset) supports this interpretation. If the sublevel splitting in this nanocrystal were instead spectrally unresolved, the line would be unpolarized. From the expected three orthogonal dipoles, we calculated the relative intensity of the photoluminescence peaks and their polarization for arbitrary observation directions (Supplementary Section 4). We then determined (Fig. 3d-f) the nanocrystal orientations consistent with the spectra and polarizations in Fig. 3a-c. Again, good agreement is obtained.

Figure 3g presents the experimental statistics for one-, two-, and three-line spectra. One is most common, suggesting that only the lowest sublevel is populated. For the two- and three-line spectra, the measured energy splittings are plotted in Fig. 3h,i. Given three sublevels separated by energies $\Delta_1$ and $\Delta_2$ (inset, Fig. 3i), the average splitting $\overline{\Delta}$ is $0.5\left(\overline{\Delta}_1 + \overline{\Delta}_2\right)$, bars denoting averages. However, two-line spectra can involve any two of the three features, leading to the average $\overline{\Delta}_1/3 + \overline{\Delta}_2/3 + \left(\overline{\Delta}_1 + \overline{\Delta}_2\right)/3 = 2\left(\overline{\Delta}_1 + \overline{\Delta}_2\right)/3$. Thus, we predict a ratio of 1.33 for average measured splittings in two- versus three-line spectra. The experimental ratio of 1.42±0.12 again supports our model.

While we have used cryogenic temperatures to confirm the bright triplet exciton, it remains important at elevated temperatures. This explains why room-temperature emission from $CsPbX_3$ perovskite nanocrystals is 20x faster than previously studied semiconductor nanocrystals[13-16]. The emission should be even faster for perovskite nanowires and nanoplatelets. Such shapes can further decrease the radiative lifetime due to diminished dielectric screening and smaller one- or two-dimensional excitons[38].

Bright-triplet emission can potentially boost efficiency in optical devices with electrical injection and improve quantum optical sources and sensors. The discovery that the lowest exciton is bright in lead halide perovskites also reveals criteria for this phenomenon. Potential semiconductors should lack inversion symmetry, and one band edge should have *s* symmetry and the other *p*, with the latter affected by strong spin–orbit coupling such that $J_{e,h}$=1/2. Finally, the Rashba coefficient for both bands must be nonzero with the same sign.

**Online Content** Methods, along with any Extended Data display items and Supplementary Information, are available in the online version of the paper; references unique to these sections appear only in the online paper.

## References

1. Scholes, G. D. & Rumbles, G. Excitons in nanoscale systems. *Nat. Mater.* **5**, 683-696 (2006).
2. Hund, F. Concerning the interpretation of complex spectra, especially the elements scandium to nickel. *Z. Phys.* **33**, 345-371 (1925).
3. Uoyama, H., Goushi, K., Shizu, K., Nomura, H. & Adachi, C. Highly efficient organic light-emitting diodes from delayed fluorescence. *Nature* **492**, 234-238 (2012).
4. Thompson, N. J., Wilson, M. W. B., Congreve, D. N., Brown, P. R., Scherer, J. M., Bischof, T. S., Wu, M. F., Geva, N., Welborn, M., Van Voorhis, T., Bulovic, V., Bawendi, M. G. & Baldo, M. A. Energy harvesting of non-emissive triplet excitons in tetracene by emissive PbS nanocrystals. *Nat. Mater.* **13**, 1039-1043 (2014).
5. Mongin, C., Garakyaraghi, S., Razgoniaeva, N., Zamkov, M. & Castellano, F. N. Direct observation of triplet energy transfer from semiconductor nanocrystals. *Science* **351**, 369-372 (2016).
6. Bychkov, Yu. A. & Rashba, E. I. Oscillatory effects and the magnetic susceptibility of carriers in inversion layers. *J. Phys. C: Solid State Phys.* **17**, 6039-6045 (1984).
7. Kim, M., Im, J., Freeman, A. J., Ihm, J. & Jin, H. Switchable S = 1/2 and J = 1/2 Rashba bands in ferroelectric halide perovskites. *Proc. Natl. Acad. Sci.* **111**, 6900-6904 (2014).
8. Zheng, F., Tan, L. Z., Liu, S. & Rappe, A. M. Rashba spin–orbit coupling enhanced carrier lifetime in $CH_3NH_3PbI_3$. *Nano Lett.* **15**, 7794-7800 (2015).
9. Kepenekian, M., Robles, R., Katan, C., Sapori, D., Pedesseau, L. & Even, J. Rashba and Dresselhaus effects in hybrid organic–inorganic perovskites: From basics to devices. *ACS Nano* **9**, 11557-11567 (2015).
10. Mosconi, E., Etienne, T. & De Angelis, F. Rashba band splitting in organohalide lead perovskites: Bulk and surface effects. *J. Phys. Chem. Lett.* **8**, 2247-2252 (2017).
11. Protesescu, L., Yakunin, S., Bodnarchuk, M. I., Krieg, F., Caputo, R., Hendon, C. H., Yang, R. X., Walsh, A. & Kovalenko, M. V. Nanocrystals of cesium lead halide perovskites ($CsPbX_3$, X = Cl, Br, and I): Novel optoelectronic materials showing bright emission with wide color gamut. *Nano Lett.* **15**, 3692-3696 (2015).
12. Rainò, G., Nedelcu, G., Protesescu, L., Bodnarchuk, M. I., Kovalenko, M. V., Mahrt, R. F. & Stöferle, T. Single cesium lead halide perovskite nanocrystals at low temperature: Fast single-photon emission, reduced blinking, and exciton fine structure. *ACS Nano* **10**, 2485-2490 (2016).
13. Crooker, S. A., Barrick, T., Hollingsworth, J. A. & Klimov, V. I. Multiple temperature regimes of radiative decay in CdSe nanocrystal quantum dots: Intrinsic limits to the dark-exciton lifetime. *Appl. Phys. Lett.* **82**, 2793-2795 (2003).
14. Wuister, S. F., van Houselt, A., de Mello Donegá, C., Vanmaekelbergh, D. & Meijerink, A. Temperature antiquenching of the luminescence from capped CdSe quantum dots. *Angew. Chem.* **43**, 3029-3033 (2004).
15. Du, H., Chen, C., Krishnan, R., Krauss, T. D., Harbold, J. M., Wise, F. W., Thomas, M. G. & Silcox, J. Optical properties of colloidal PbSe nanocrystals. *Nano Lett.* **2**, 1321-1324 (2002).
16. Bischof, T. S., Correa, R. E., Rosenberg, D., Dauler, E. A. & Bawendi, M. G. Measurement of emission lifetime dynamics and biexciton emission quantum yield of individual InAs colloidal nanocrystals. *Nano Lett.* **14**, 6787-6791 (2014).
17. Nirmal, M., Norris, D. J., Kuno, M., Bawendi, M. G., Efros, Al. L. & Rosen, M. Observation of the 'dark exciton' in CdSe quantum dots. *Phys. Rev. Lett.* **75**, 3728-3731 (1995).

18. Klimov, V. I., *Nanocrystal Quantum Dots*, 2nd ed. (CRC Press, Boca Raton, 2010).

19. Shirasaki, Y., Supran, G. J., Bawendi, M. G. & Bulovic, V. Emergence of colloidal quantum-dot light-emitting technologies. *Nature Photon.* **7**, 13-23 (2013).

20. Dai, X., Zhang, Z., Jin, Y., Niu, Y., Cao, H., Liang, X., Chen, L., Wang, J. & Peng, X. Solution-processed, high-performance light-emitting diodes based on quantum dots. *Nature* **515**, 96-99 (2014).

21. Klimov, V. I., Mikhailovsky, A. A., Xu, Su, Malko, A., Hollingsworth, J. A., Leatherdale, C. A., Eisler, H.-J. & Bawendi, M. G. Optical gain and stimulated emission in nanocrystal quantum dots. *Science* **290**, 314-317 (2000).

22. Dang, C., Lee, J., Breen, C., Steckel, J. S., Coe-Sullivan, S. & Nurmikko, A. Red, green and blue lasing enabled by single-exciton gain in colloidal quantum dot films. *Nat. Nanotechnol.* **7**, 335-339 (2012).

23. Kim, T. H., Cho, K. S., Lee, E. K., Lee, S. J., Chae, J., Kim, J. W., Kim, D. H., Kwon, J. Y., Amaratunga, G., Lee, S. Y., Choi, B. L., Kuk, Y., Kim, J. M. & Kim, K. Full-colour quantum dot displays fabricated by transfer printing. *Nature Photon.* **5**, 176-182 (2011).

24. Bir, G. L. & Pikus, G. E., *Symmetry and Strain-Induced Effects in Semiconductors* (Wiley, New York, 1974).

25. Even, J., Pedesseau, L., Jancu, J.-M. & Katan, C. Importance of spin–orbit coupling in hybrid organic/inorganic perovskites for photovoltaic applications. *J. Phys. Chem. Lett.* **4**, 2999-3005 (2013).

26. Huang, L.-Y. & Lambrecht, W. R. L. Electronic band structure, phonons, and exciton binding energies of halide perovskites $CsSnCl_3$, $CsSnBr_3$, and $CsSnI_3$. *Phys. Rev. B* **88**, 165203 (2013).

27. Kane, E. O. in *Semiconductors and Semimetals*, edited by R. K. Willardson and A. C. Beer (Academic Press, New York, 1966), Vol. 1, pp. 75-100.

28. Yu, P. Y. & Cardona, M., *Fundamentals of Semiconductors* (Springer-Verlag, Berlin, 2001).

29. Koster, G. F., Dimmock, J. O., Wheeler, R. G. & Statz, H., *Properties of the Thirty-Two Point Groups* (MIT Press, Cambridge, 1963).

30. Yaffe, O., Guo, Y., Tan, L. Z., Egger, D. A., Hull, T., Stoumpos, C. C., Zheng, F., Heinz, T. F., Kronik, L., Owen, J. S., Rappe, A. M., Pimenta, M. A. & Brus, L. E. Local polar fluctuations in lead halide perovskite crystals. *Phys. Rev. Lett.* **118**, 136001 (2017).

31. Galland, C., Ghosh, Y., Steinbruck, A., Sykora, M., Hollingsworth, J. A., Klimov, V. I. & Htoon, H. Two types of luminescence blinking revealed by spectroelectrochemistry of single quantum dots. *Nature* **479**, 203-207 (2011).

32. Edwards, T. W. & Van Bladel, J. Electrostatic dipole moment of a dielectric cube. *Appl. Sci. Res. B* **9**, 151-155 (1961).

33. Cottingham, P. & Brutchey, R. L. On the crystal structure of colloidally prepared $CsPbBr_3$ quantum dots. *Chem. Commun.* **52**, 5246-5249 (2016).

34. Landau, L. D. & Lifshitz, E. M., *Electrodynamics of Continuous Media* (Pergamon Press, Oxford, 1960).

35. Efros, Al. L. & Efros, A. L. Interband absorption of light in a semiconductor sphere. *Sov. Phys. Semicond.* **16**, 772-775 (1982).

36. Rashba, E. I. & Gurgenishvili, G. E. Edge absorption theory in semiconductors. *Sov. Phys. Solid State* **4**, 759-760 (1962).

37. Fu, N.; Tamarat, P.; Huang, H.; Even, J.: Rogach, A. L.; Lounis, B. Neutral and charged exciton fine structure in single lead halide perovskite nanocrystals revealed by magneto-optical spectroscopy. *Nano Lett.* **17**, 2895-2901 (2017).

38. Rodina, A. V. & Efros, Al. L. Effect of dielectric confinement on optical properties of colloidal nanostructures. *J. Exp. Theor. Phys.* **122**, 554-566 (2016).

**Acknowledgements** We thank F. Krieg for providing CsPbBr$_3$ nanocrystals. Al.L.E. thanks E. Ivchenko, M. Glazov, and E. Rashba for useful discussions. M.A.B., G.R., T.S., M.V.K., and R.F.M. acknowledge the European Union's Horizon-2020 programme through the Marie-Sklodowska Curie ITN network PHONSI (H2020-MSCA-ITN-642656) and the Swiss State Secretariat for Education Research and Innovation (SERI). J.G.M., S.G.L., N.B., J.L.L., and Al.L.E. acknowledge support from the U.S. Office of Naval Research (ONR) through the core funding of the Naval Research Laboratory. R.V. was funded by ONR Grant N0001416WX01849. A.S. acknowledges support from the Center for Advanced Solar Photophysics (CASP), an Energy Frontier Research Center (EFRC) funded by BES, OS, U.S. DOE. D.J.N. and M.V.K. acknowledge partial financial support from the European Research Council under the European Union's Seventh Framework Program (FP/2007-2013) / ERC Grant Agreement Nr. 339905 (QuaDoPS Advanced Grant) and Nr. 306733 (NANOSOLID Starting Grant), respectively.

**Author contributions** This work resulted from ongoing interactions between G.R., M.V.K., D.J.N., and Al.L.E. M.A.B., G.R., and T.S. performed the optical experiments. They analyzed and interpreted the data with input from D.J.N., R.F.M., P.C.S., and Al.L.E. Al.L.E. conceived the model and supervised the theoretical research. R.V. calculated the radiative lifetimes and developed theory for the observed exciton fine structure. R.V. and A.S. developed the 4-band model describing the energy dispersion at the R-point and calculated the fine structure using the wavefunction extracted from first-principle calculations. P.C.S. developed the anisotropic exchange Hamiltonian and performed the group-theory analysis of the fine structure and selection rules. M.J.M., N.B., and J.L.L. completed the first-principle calculations of the bulk band structures and the band-edge wavefunctions. R.V., P.C.S, and Al.L.E. developed the effective exchange and Rashba Hamiltonian describing the exciton fine structure. J.G.M. and S.G.L. calculated the internal electric fields in spherical and cube-shaped nanocrystals. G.N. prepared the samples and performed electron microscopy under the supervision of M.V.K. Al.L.E. and D.J.N. wrote the manuscript with input from all authors.

**Author Information** The authors declare no competing financial interests. Readers are welcome to comment on the online version of the paper. Correspondence and requests for materials should be addressed to M.V.K. (mvkovalenko@ethz.ch), G.R. (rainog@ethz.ch), D.J.N. (dnorris@ethz.ch), or Al.L.E. (sasha.efros@nrl.navy.mil).

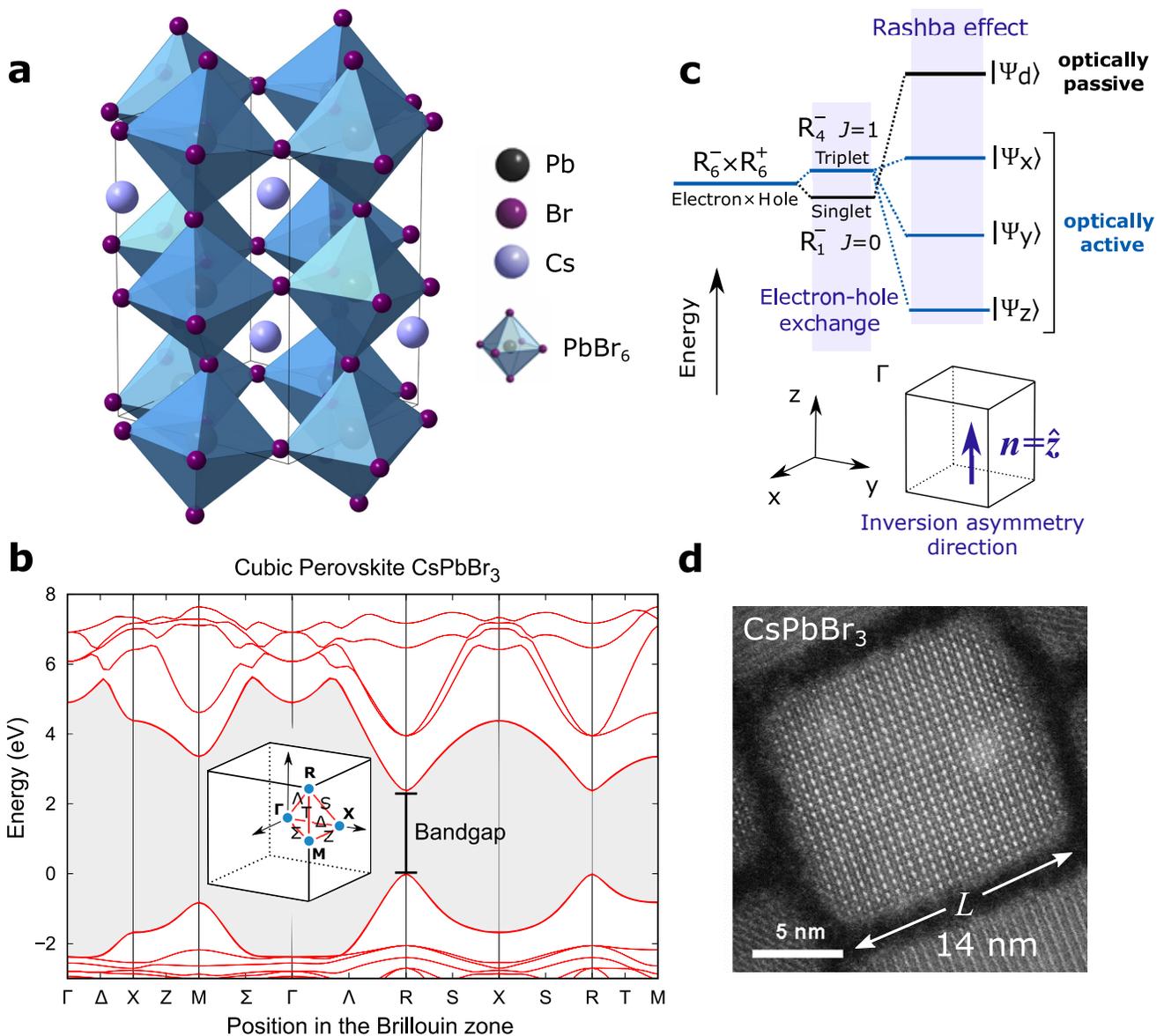

**Figure 1 | Crystal and electronic structure for perovskite CsPbBr₃. a**, Unit cell for perovskite CsPbBr₃. The structure is quasi-cubic with an orthorhombic distortion. **b**, Calculated band structure of cubic perovskite CsPbBr₃. The inset shows the first Brillouin zone of the cubic crystal lattice. **c**, The expected fine structure of the band-edge exciton considering short-range electron–hole exchange (middle) and then including the Rashba effect (right) under orthorhombic symmetry. The latter splits the exciton into three bright states with transition dipoles oriented along the orthorhombic symmetry axes (labelled *x*, *y*, and *z*) and a higher energy dark state. The energetic order of the three lower sublevels is determined by the orthorhombic distortion. The orthorhombic unit cell and the resulting sublevel order is shown for CsPbBr₃. **d**, Transmission electron micrograph of an individual CsPbBr₃ nanocrystal of edge length *L*=14 nm.

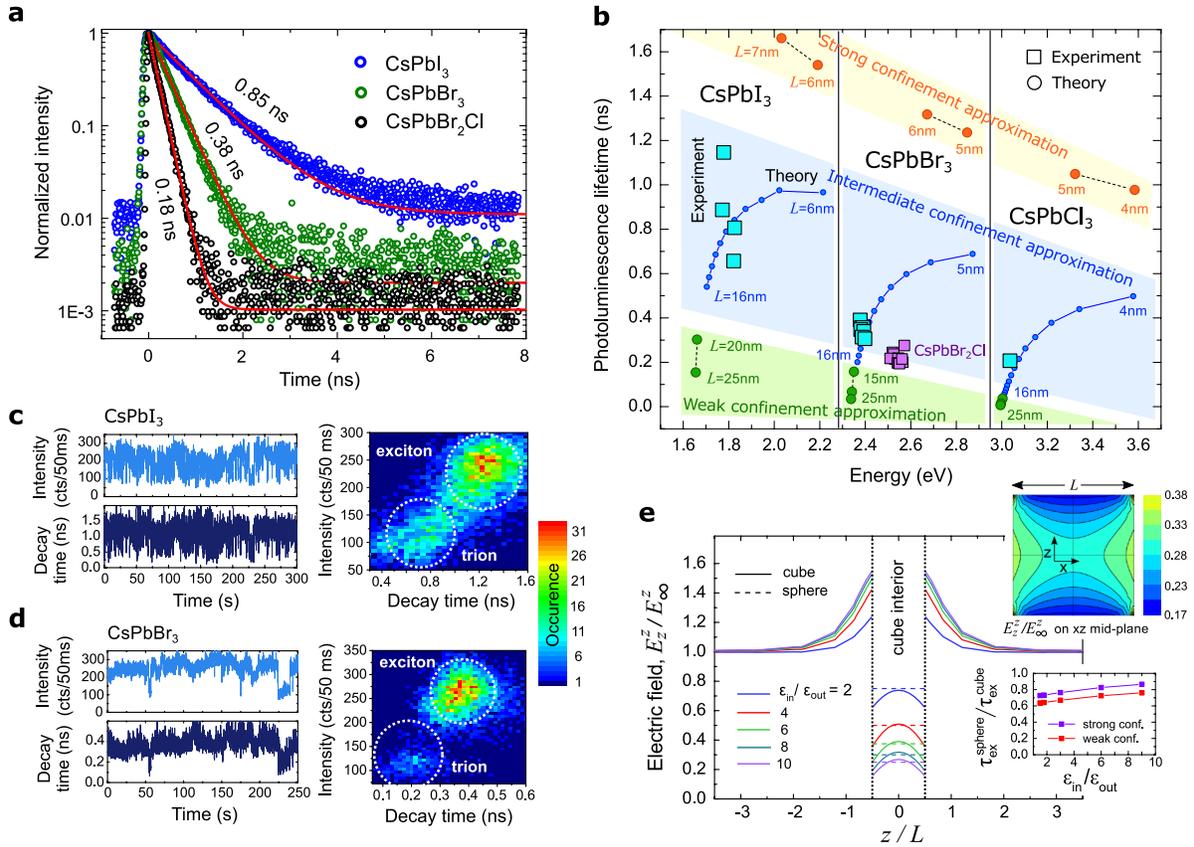

**Figure 2 | Characterization of fast radiative lifetimes in CsPbX₃ nanocrystals. a**, Photoluminescence decays (5 K) measured from single CsPbI₃ (*L*=14 nm), CsPbBr₃ (11 nm), and CsPbBr₂Cl (14 nm) perovskite nanocrystals. **b**, Calculated radiative lifetimes of the bright triplet exciton versus transition energy for CsPbX₃ nanocrystals with X=Cl, Br, and I. The theoretical results are divided into three size regimes: strong (orange circles), intermediate (blue circles), and weak (green circles) exciton confinement. These values are compared with measured photoluminescence decays (5 K) from individual perovskite nanocrystals (squares, sizes as in **a**). A data point for an ensemble of CsPbCl₃ nanocrystals (*L*=10 nm) is also shown. Measured values are consistent with calculations in the intermediate confinement regime, which include electron–hole correlations. **c,d** Detected photon counts (left panels) versus time (5 K) from individual CsPbI₃ and CsPbBr₃ nanocrystals (sizes as in **a**). Traces show "A-type" blinking from the nanocrystals[31]. Such data can be analyzed to separate contributions to the photoluminescence decay from exciton and trion emission (right panel). **e**, Calculated distribution of the *z* component of the electric field, $E_z^z$, normalized to the applied field (along the *z* direction) at infinite distance, $E_\infty^z$, *i.e.* $E_z^z / E_\infty^z$. This quantity is plotted versus position *z* across the center line of spherical (dashed lines) or cube-shaped (solid lines) nanocrystals for various ratios of the dielectric constant inside ($\varepsilon_{in}$) to outside ($\varepsilon_{out}$) the nanocrystal. The field inside the nanocrystal is essentially always lower for the cube compared to the sphere. Upper inset: Calculated two-dimensional distribution of $E_z^z / E_\infty^z$ inside a cube-shaped nanocrystal plotted on the *xz* mid-plane. The ratio $\varepsilon_{in}/\varepsilon_{out}$ was 6. Lower inset: Calculated ratio of radiative decay times for spherical and cubical nanocrystals with the same volume versus $\varepsilon_{in}/\varepsilon_{out}$ for strong and weak confinement.

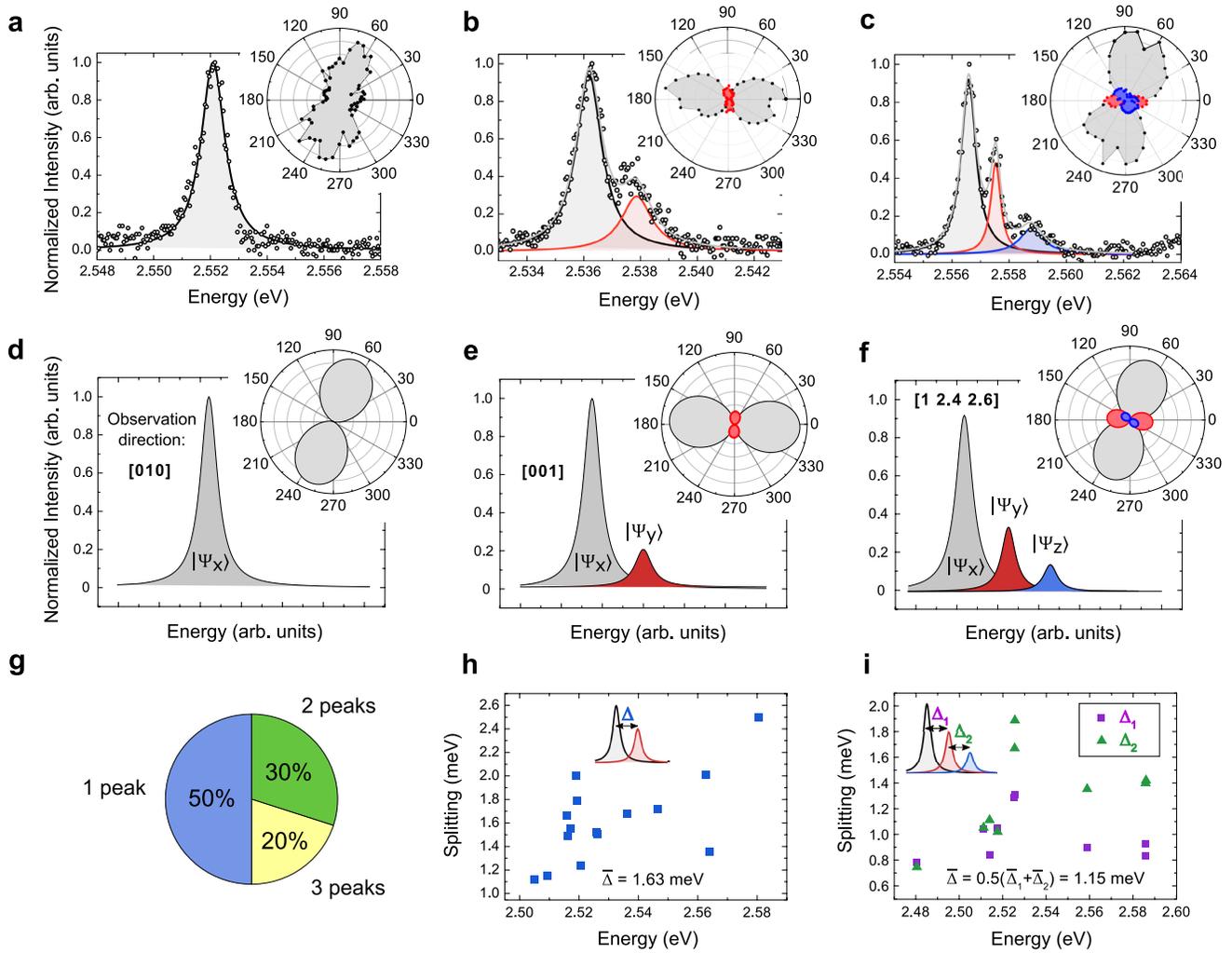

**Figure 3 | Fine structure of the bright triplet exciton for CsPbBr₂Cl nanocrystals. a-c**, Photoluminescence spectra (5 K) of individual nanocrystals (*L*=14±1 nm) exhibiting a single peak, two peaks, and three peaks. The inset shows the polarization of each of the spectral features. For the spectra, a linear polarizer was placed in the detection path. The angle of this polarizer was adjusted such that the relative intensity of the features in the spectra matched the polarization dependence in the insets. **d-f**, Simulated spectra and polarizations for nanocrystal orientations that match the experimental results in **a-c**. See Supplementary Section 4 for details. Each panel lists the observation direction required relative to the orthorhombic unit-cell axes. **g**, Experimental statistics for observation of single-peak, two-peak, and three-peak spectra from individual nanocrystals with *L*=7.5-14 nm (51 spectra with 35 splittings in total). **h,i** Experimental fine-structure splitting measured for the two-peak and three-peak spectra, respectively. The average splitting in each case is provided.

## Methods

**Chemicals.**

The following reagents were used to prepare CsPbX$_3$ nanocrystals: cesium carbonate (Cs$_2$CO$_3$, Aldrich, 99.9%), 1-octadecene (ODE, Sigma-Aldrich, 90%), oleic acid (OA, Sigma-Aldrich, 90%), oleylamine (OAm, Acros Organics, 80-90%), lead chloride (PbCl$_2$, ABCR, 99.999%), lead bromide (PbBr$_2$, ABCR, 98%), lead iodide (PbI$_2$, ABCR, 99.999%), n-trioctylphosphine (TOP, Strem, 97%), hexane (Sigma-Aldrich, ≥95%), and toluene (Fischer Scientific, HPLC grade).

**Synthesis.**

The CsPbX$_3$ (X=Cl, Br, and I) and CsPbBr$_2$Cl nanocrystals were synthesized by fast reaction between Cs-oleate and PbX$_2$ in the presence of OA and OAm (TOP is also added for CsPbCl$_3$ and CsPbBr$_2$Cl nanocrystals). First, the Cs-oleate was prepared by loading Cs$_2$CO$_3$ (0.407 g) into a 50-ml 3-neck flask along with ODE (20 ml) and OA (1.25 ml). The mixture is dried under vacuum for 1 h at 120 °C and then switched to N$_2$. Since Cs-oleate precipitates out of ODE at room temperature, it must be pre-heated to 100 °C before injection. The ODE, OA, and OAm were pre-dried before use by degassing under vacuum at 120 °C for 1 h. For the nanocrystal reaction, 0.376 mmol PbX$_2$ (X=Cl, Br, or I), dried OA (3 ml for PbCl$_2$,1 ml for PbBr$_2$, or 1.5 ml for PbI$_2$), dried OAm (3 ml for PbCl$_2$, 1 ml for PbBr$_2$, or 1.5 ml for PbI$_2$), and dried ODE (5 ml) were combined in a 25-ml 3-neck flask. For CsPbCl$_3$, TOP (1 ml) was also added. The mixture was then degassed for 10 min under vacuum at 120 °C, and the flask was filled with N$_2$ and heated to 200 °C. Cs-oleate (0.8 ml from the stock solution prepared as described above) was swiftly injected when 200 °C was reached. After 10 s the reaction was stopped by cooling the reaction system with a water bath. The solution was centrifuged (4 min, 13750 g) and the supernatant discarded. Hexane (0.3 ml) was added to the precipitate to disperse the nanocrystals and centrifuged again. The obtained precipitate was redispersed in 3 ml toluene and centrifuged (2 min, 2200 g). The supernatant was separated from the precipitate, filtered, and used for our investigations. For $CsPbBr_2Cl$, 0.094 mmol $PbCl_2$, 0.282 mmol $PbBr_2$, dried OA (1.5 ml), dried OAm (1.5 ml), TOP (1 ml), and dried ODE (5 ml) were loaded into a 25-ml 3-neck flask and the same protocol was followed.

**Sample preparation.**

For single-nanocrystal spectroscopy, the colloidal dispersions from the above syntheses were diluted to nanomolar concentrations in solutions of 3 mass percent polystyrene in toluene. This dispersion was then spin-casted at 5000 r.p.m. onto intrinsic crystalline Si wafers with a 3-µm-thick thermal-oxide layer. For ensemble measurements, the undiluted nanocrystal dispersions from the previous section were drop-casted on glass substrates.

**Optical characterization.**

All optical measurements of single nanocrystals were performed in a self-built micro-photoluminescence (µ-PL) setup. The samples were mounted on xyz nano-positioning stages inside an evacuated liquid-helium flow cryostat and cooled down to 5 K. Single nanocrystals were excited by means of a fiber-coupled excitation laser at an energy of 3.06 eV with a repetition rate of 40 MHz and a pulse duration of 50 ps. The excitation beam was sent through a linear polarizer and a short-wavelength-pass filter before being directed toward the sample by a dichroic beam splitter. Typical power densities used to excite single nanocrystals were 2-120 W/cm$^2$. For both excitation and detection, a long-working distance 100x microscope objective with numerical aperture of 0.7 was used. The nearly Gaussian excitation spot had a 1/e$^2$ diameter of 1.4 µm. The emission was filtered using a long-pass filter and dispersed by a 0.75 m monochromator with an 1800 lines/mm grating before detection with a back-illuminated, cooled CCD camera. For polarization-dependent measurements, a liquid crystal retarder was employed to compensate for retardation effects in the setup. For photoluminescence lifetime and time-tagged time-resolved (TTTR3) single-photon-counting measurements, we filtered the emission with a suitable tunable bandpass filter to either measure only the excitonic photoluminescence decay or to correlate excitonic and trionic emission intensities and decay times with a time-correlated single-photon-counting system with nominal time resolution of 30 ps.

Ensemble measurements were performed in an exchange-gas cryostat at 5 K. Here, the samples were excited with a frequency-doubled Ti:sapphire femtosecond pulsed laser with a repetition rate of 80 MHz at 3.1 eV. The emitted light was dispersed by a 150 lines/mm grating within a 300-mm focal length spectrograph and detected by a streak camera with 2 ps resolution.

**Band-structure calculations.**

Figure 1b and Extended Data Fig. 1 show calculated band structures for $CsPbBr_3$, $CsPbCl_3$, and $CsPbl_3$. We assume that these materials exist in the cubic perovskite structure with a lattice constant of 5.865, 5.610, and 6.238 Å, respectively[39]. The electronic structure of these crystals was determined using the Vienna *Ab-initio* Simulation Package (VASP)[40-42] with projector-augmented wavefunctions[43]. Our initial calculations used the PBEsol[44,45] generalized gradient approximation, and included spin–orbit coupling. We used an energy cutoff of 400 eV and Γ-centered k-point grid of 6×6×6, which yield 40 k-points in the irreducible Brillouin zone.

As expected, standard density functional theory (DFT) seriously underestimates the bandgap in these materials. Accordingly, we used a modified version of the Heyd-Scuseria-Ernzerhof "HSE06" hybrid functional[46], which mixes exact Hartree-Fock exchange with conventional DFT. We initially started with 25% mixing, and planned to adjust the mixing to match the observed bandgap. However, this was not possible, even with 45% Hartree-Fock in the calculation for $CsPbBr_3$. This produced a bandgap of 1.4 eV, far smaller than the experimental gap of 2.8 eV. Rather than using even higher mixing, or even a full-scale Hartree-Fock calculation, we instead added a scissors operator to adjust the bandgap to the experimental result. We found that the electron and hole masses were nearly unchanged with Hartree-Fock mixing, leading us to believe that this technique still provides the correct physics. Further confirmation was provided by conducting $G_0W_0$ calculations (also with VASP) on top of the PBE results. For this approach, we employed a plane-wave energy cutoff of 600 eV, a 150 eV energy cutoff for the response functions, 1894 unoccupied states, spin–orbit coupling, and "GW" pseudopotentials including all semicore electrons. Although these calculations yielded band gaps that were in closer agreement with experiment (1.96 eV for $CsPbI_3$, 2.36 eV for $CsPbBr_3$, and 3.27 eV for $CsPbCl_3$), other aspects of the band structure remained virtually unchanged.

## Method References

39. Mehl, M. J., Hicks, D., Toher, C., Levy, O., Hanson, R. M., Hart, G. & Curtarolo, S. The AFLOW library of crystallographic prototypes. Preprint at https://arxiv.org/abs/1607.02532 (2016).
40. Kresse, G., *Ab Initio* Molekular Dynamik für Flüssige Metalle, Ph.D. thesis, Technische Universität Wien, 1993.
41. Kresse, G. & Hafner, J. *Ab initio* molecular dynamics for open-shell transition metals. *Phys. Rev. B* **48**, 13115-13118 (1993).
42. Kresse, G. & Hafner, J. *Ab initio* molecular-dynamics simulation of the liquid-metal–amorphous-semiconductor transition in germanium. *Phys. Rev. B* **49**, 14251-14269 (1994).
43. Blöchl, P. E. Projector augmented-wave method. *Phys. Rev. B* **50**, 17953-17979 (1994).
44. Perdew, J. P., Ruzsinszky, A., Csonka, G. I., Vydrov, O. A., Scuseria, G. E., Constantin, L. A., Zhou, X. & Burke, K. Restoring the density-gradient expansion for exchange in solids and surfaces. *Phys. Rev. Lett.* **100**, 136406 (2008).
45. Perdew, J. P., Ruzsinszky, A., Csonka, G. I., Vydrov, O. A., Scuseria, G. E., Constantin, L. A., Zhou, X. & Burke, K. Erratum: Restoring the density-gradient expansion for exchange in solids and surfaces. *Phys. Rev. Lett.* **102**, 039902 (2009).

46. Krukau, A. V., Vydrov, O. A., Izmaylov, A. F. & Scuseria, G. E. Influence of the exchange screening parameter on the performance of screened hybrid functionals. *J. Chem. Phys.* **125**, 224106 (2006).

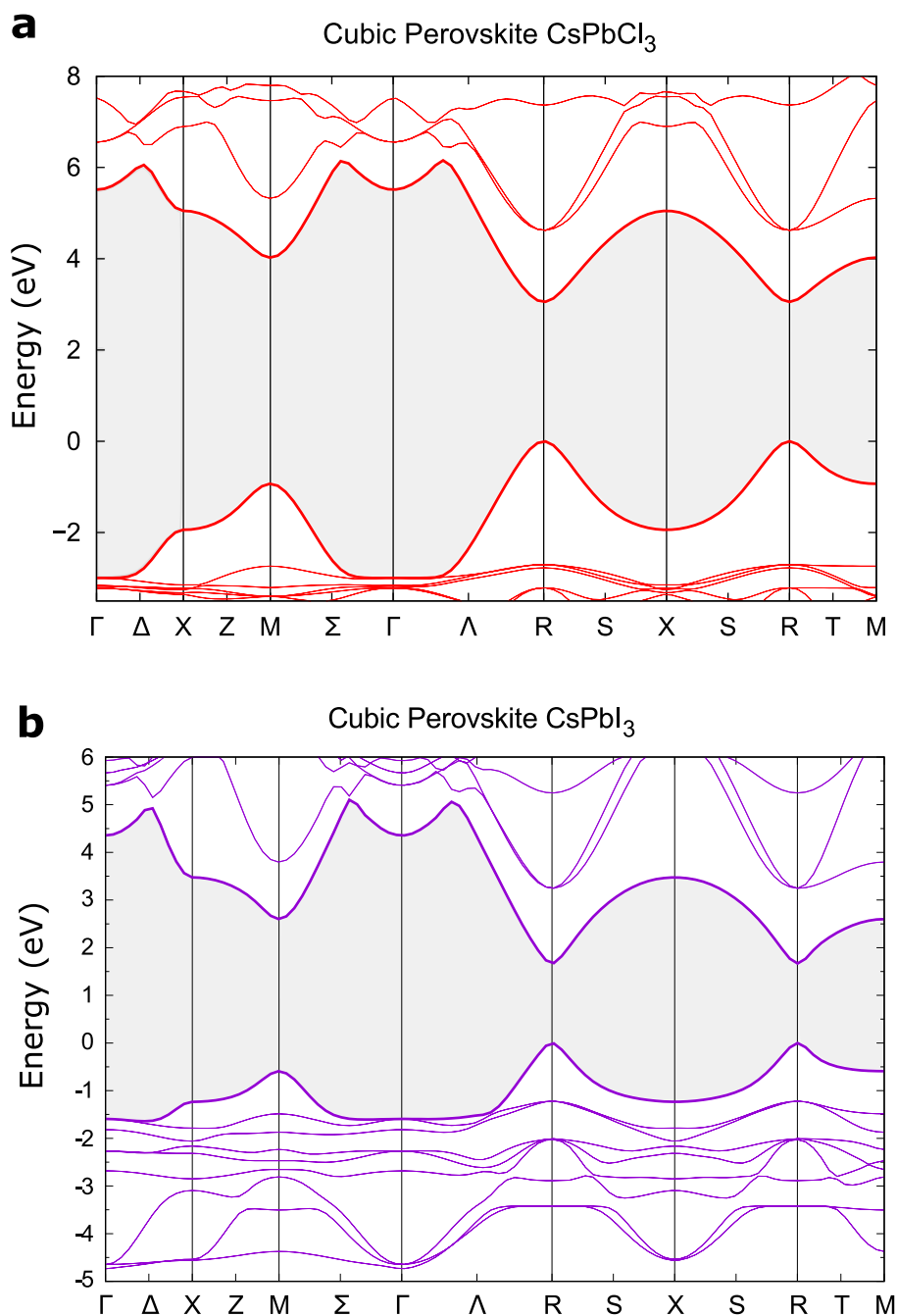

**Extended Data Figure 1 | Electronic structure for perovskite CsPbCl₃ and CsPbI₃. a**, Calculated band structure of cubic perovskite CsPbCl₃. **b**, Calculated band structure of cubic perovskite CsPbI₃. See Methods for details about the calculations.

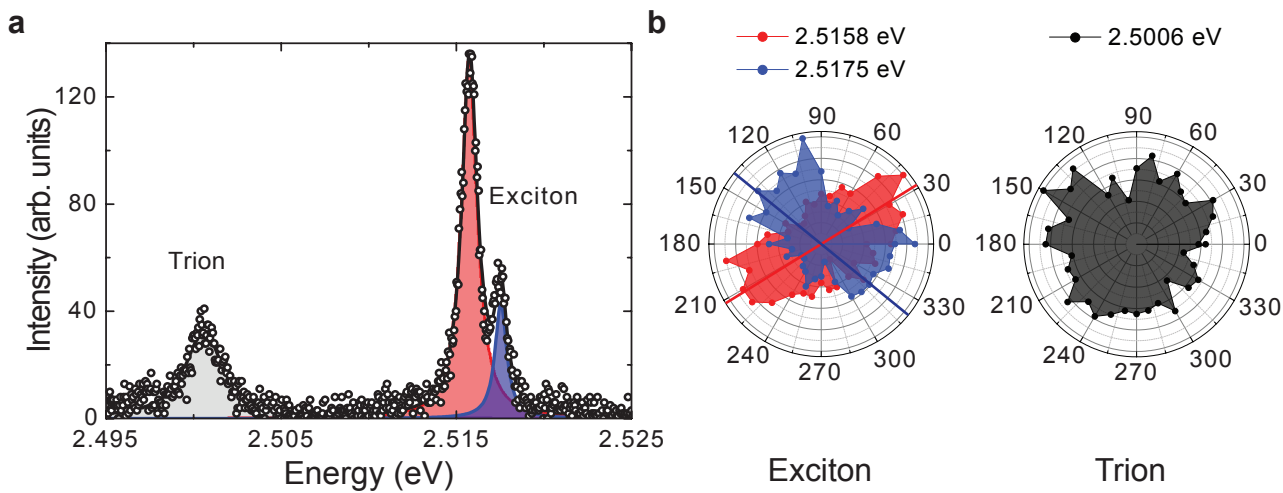

**Extended Data Figure 2 | Exciton and trion emission from an individual CsPbBr₂Cl nanocrystal. a**, Photoluminescence spectrum of a single CsPbBr₂Cl nanocrystal showing two exciton peaks at 2.5158 and 2.5175 eV and a trion peak that is red-shifted by 15-17 meV. **b**, Polarization properties of the exciton (left plot) and trion (right plot) emission peaks. The normalized area of a Lorentzian-peak fit for two exciton peaks (red and blue) and the trion peak (black) are shown as a function of the linear polarizer angle (placed in front of the spectrograph). Both exciton peaks show a dominantly linear polarization, with the main axis being indicated by the blue and red lines. The trion emission is unpolarized. See Supplementary Section 4 for further discussion.

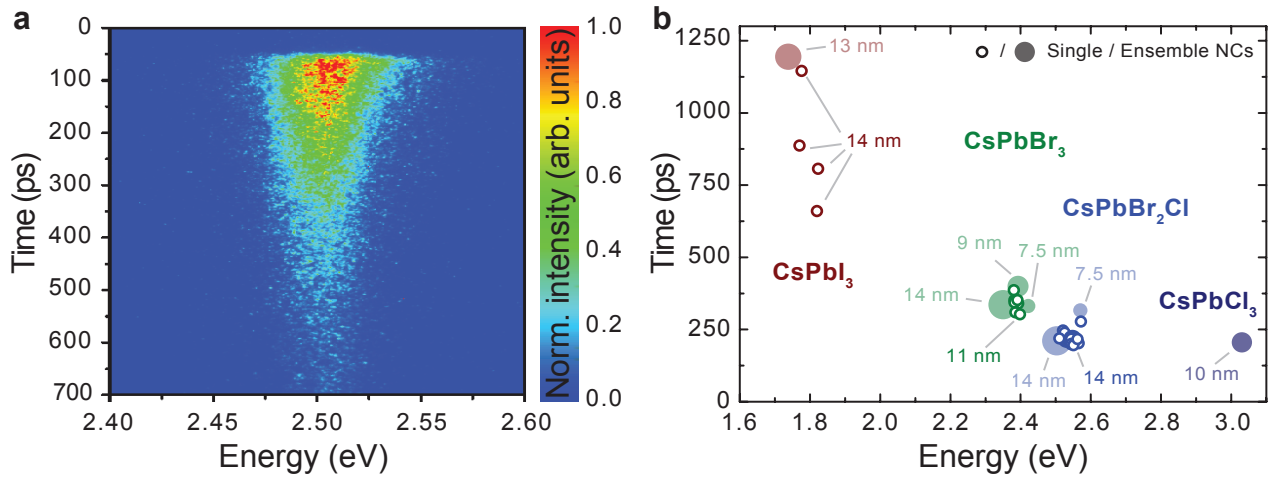

**Extended Data Figure 3 | Composition-dependent ensemble photoluminescence decay measurements of lead halide perovskite nanocrystals. a**, Typical streak-camera measurement of the photoluminescence from an ensemble of CsPbBr$_2$Cl nanocrystals at 5 K. In this example, the nanocrystals have $L$=14 nm. The emission peak is centered at 2.51 eV, and the exponential decay time is 210 ps, as extracted by summing over all energies, which is in good agreement with the results for single CsPbBr$_2$Cl nanocrystals of the same size. The ensemble decay spectrum is slightly asymmetric (being faster at higher energies), which might originate from the activation of an energy-transfer process from smaller to larger nanocrystals. To account for the latter effect, we have only considered the long component of the decay curve. **b**, Photoluminescence lifetimes at 5 K extracted for ensemble samples of nanocrystals of various compositions and sizes. The ensemble data (solid circles) are compared with single-nanocrystal measurements (open circles). The good agreement between the two data sets is further evidence that the measured single-nanocrystal photoluminescence decays are due to fast exciton radiative lifetimes and not due to trions, as the ensemble data are acquired at very low excitation power where photo-generated charging is not observed.

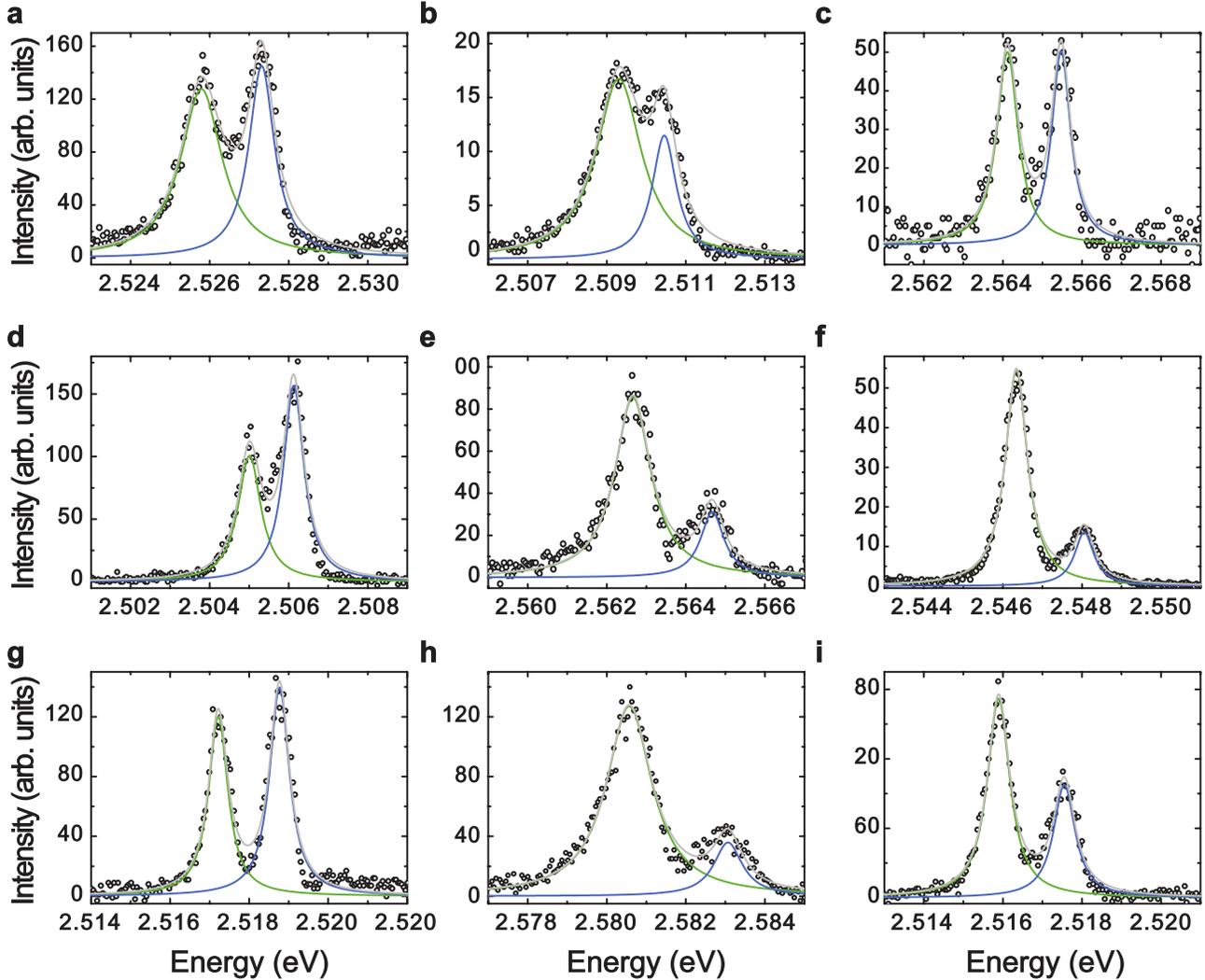

**Extended Data Figure 4 | Representative 2-peak spectra for individual CsPbBr₂Cl nanocrystals at 5 K. a-i**, Photoluminescence spectra of single nanocrystals. Each spectrum was recorded with a linear polarizer in the detection path. Thus, the displayed relative intensities cannot be used to determine the relative (potentially thermal) population within the fine structure multiplet. The linear polarizer was used here because it can be rotated to resolve all spectral features. Without the polarizer, the low-energy peak typically dominates in intensity.

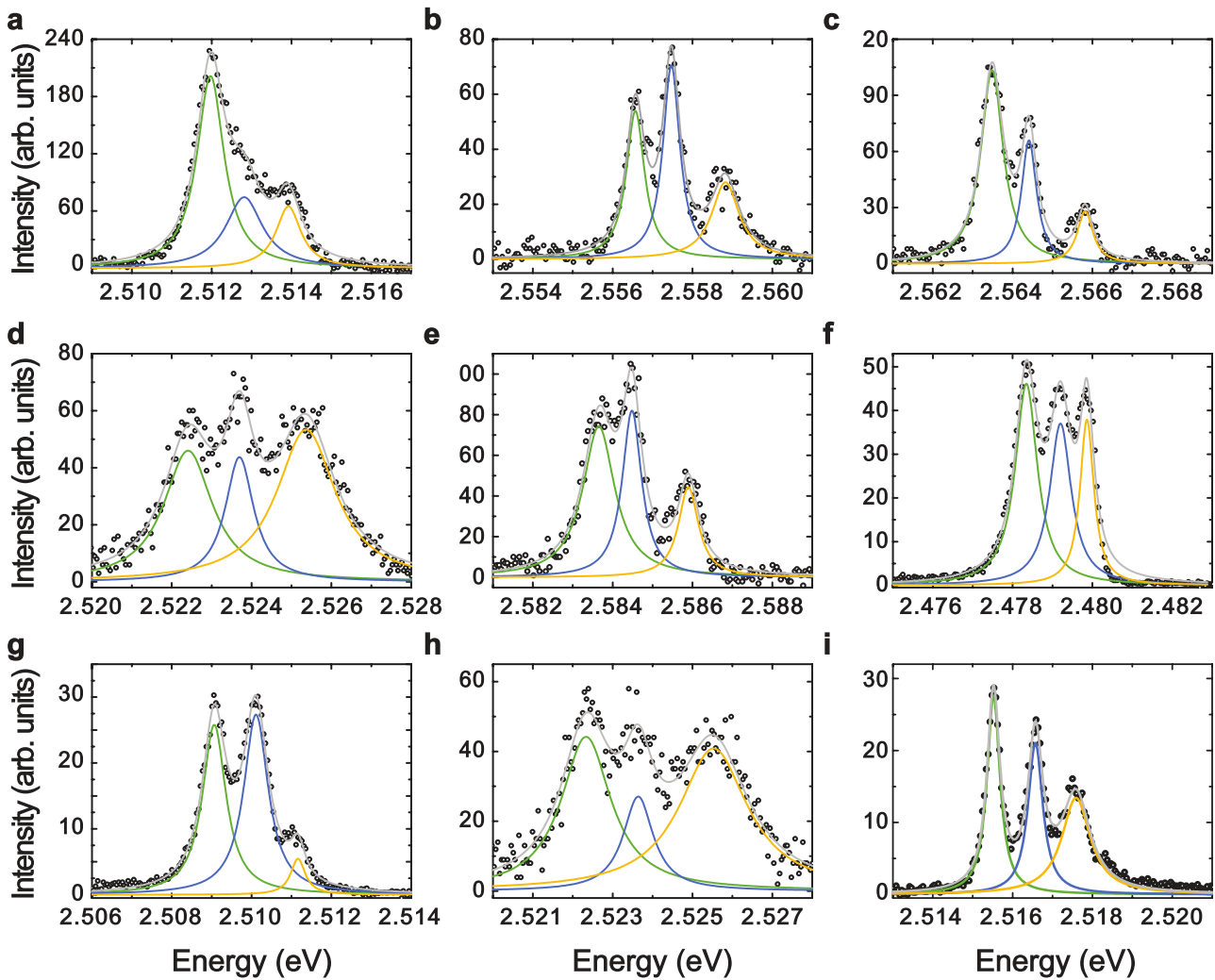

**Extended Data Figure 5 | Representative 3-peak spectra for individual CsPbBr₂Cl nanocrystals at 5 K. a-i**, Photoluminescence spectra of single nanocrystals. Each spectrum was recorded with a linear polarizer in the detection path. Thus, the displayed relative intensities cannot be used to determine the relative (potentially thermal) population within the fine structure multiplet. The linear polarizer was used here because it can be rotated to resolve all spectral features. Without the polarizer, the low-energy peak typically dominates in intensity.

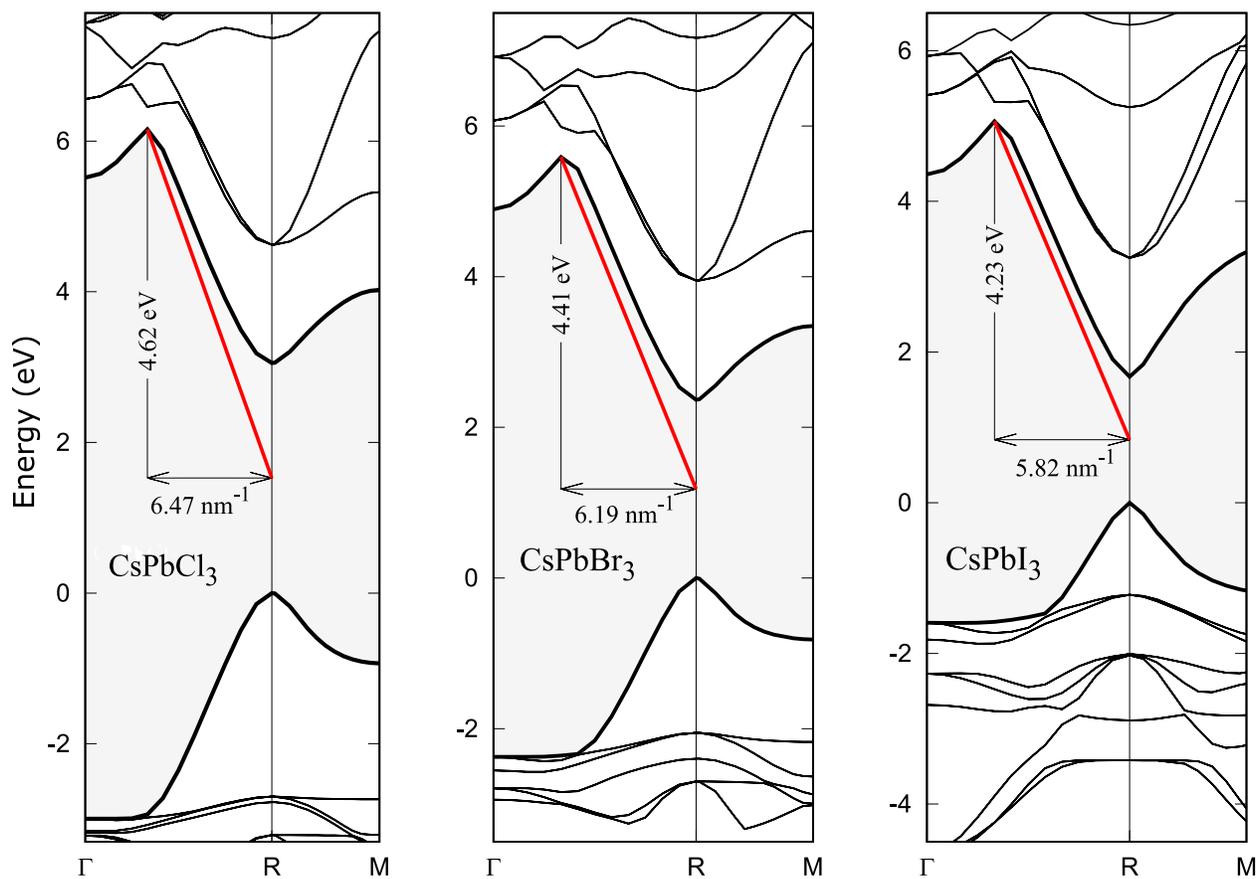

**Extended Data Figure 6 | Extraction of the Kane energy, $E_p$, for the lead halide perovskites.** From the band structures presented in Fig. 1b and Extended Data Fig. 1, the Kane energy, defined according to equation (5) in the main text, can be extracted for CsPbCl$_3$, CsPbBr$_3$, and CsPbI$_3$ from the band structure near the band edges. See Supplementary Section 1.B for details.

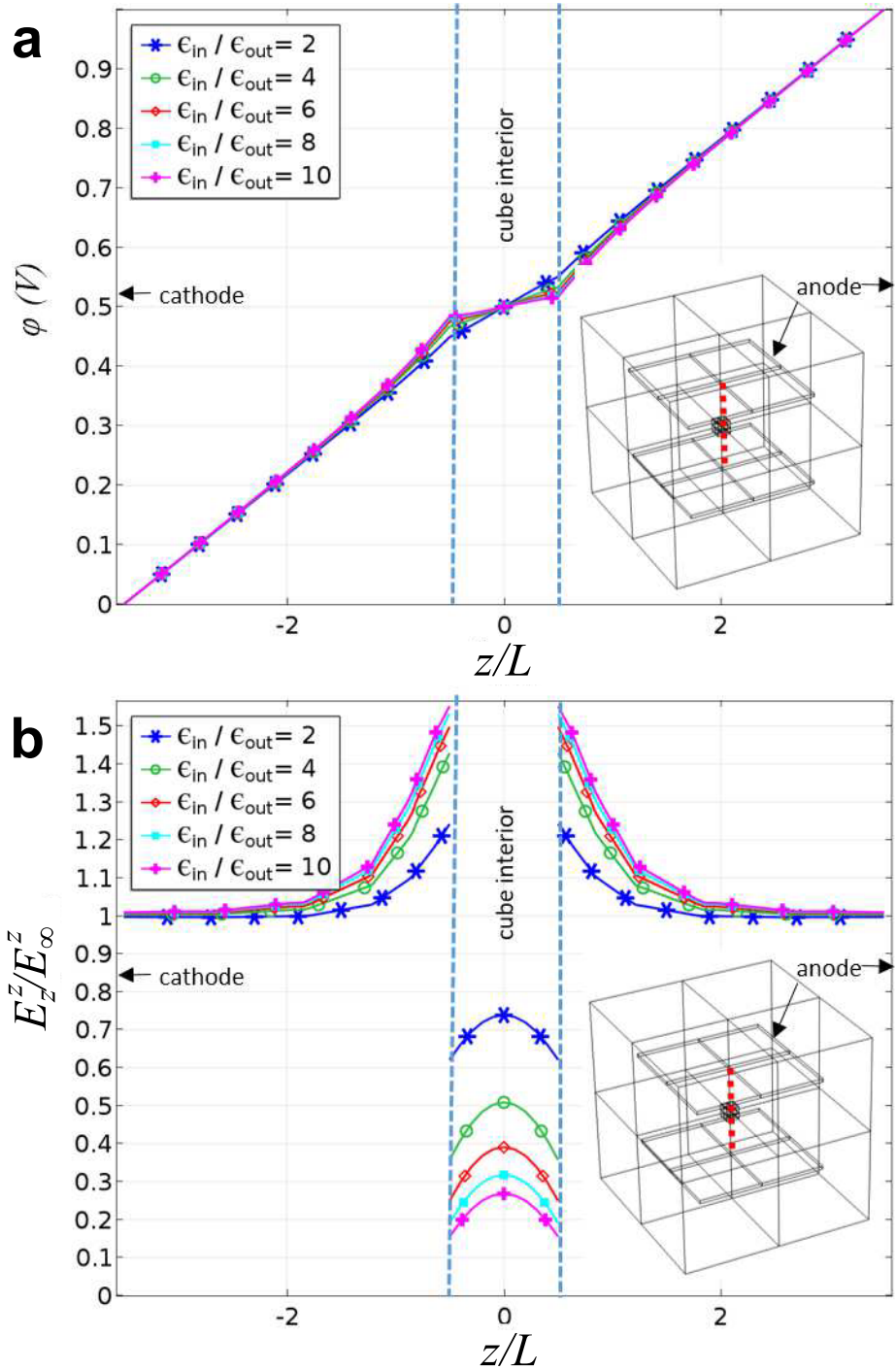

**Extended Data Figure 7 | Calculation of the interior electric field in cube-shaped nanocrystals. a**, Line plot of the electric potential, Φ, along the center line between the capacitor plates (see Supplementary Section 3.B). **b**, Line plot of the normalized electric-field magnitude, $E_z^z / E_\infty^z$, along the center line between the capacitor plates.

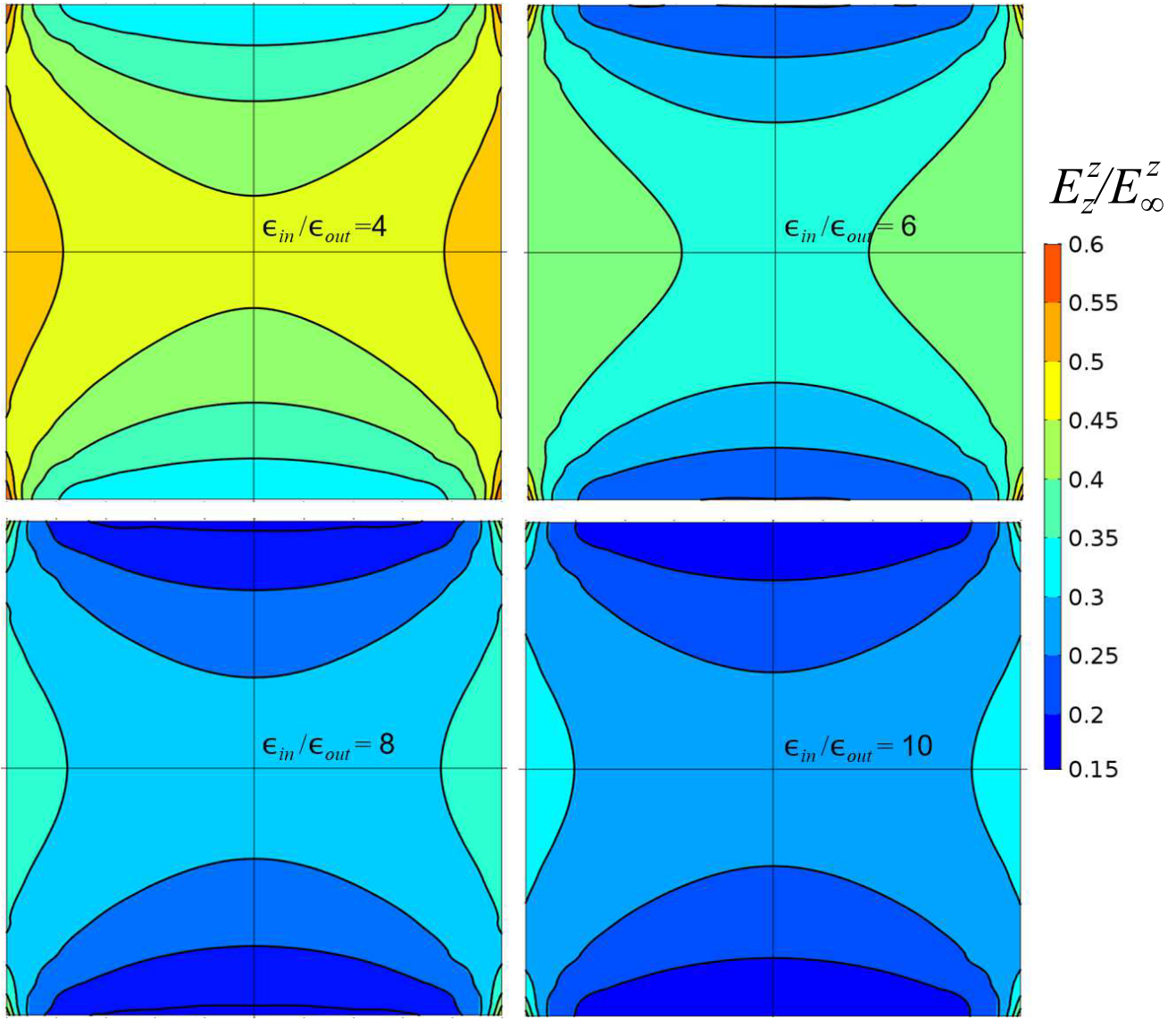

**Extended Data Figure 8 | Contour plots of normalized electric-field magnitude across a cube-shaped nanocrystal.** Contour plots of $E_z^z / E_\infty^z$ for four different ratios of the dielectric constant inside the nanocrystal ($\varepsilon_{in}$) to the surrounding medium ($\varepsilon_{out}$) (see Supplementary Section 3.B). The plots depict the *xz* mid-plane of the cube and is valid for the symmetry-equivalent *yz* mid-plane. The *z* direction is vertical. The perturbations near the corners of the plots are artifacts of the interpolation resolution utilized by the software employed to construct them.

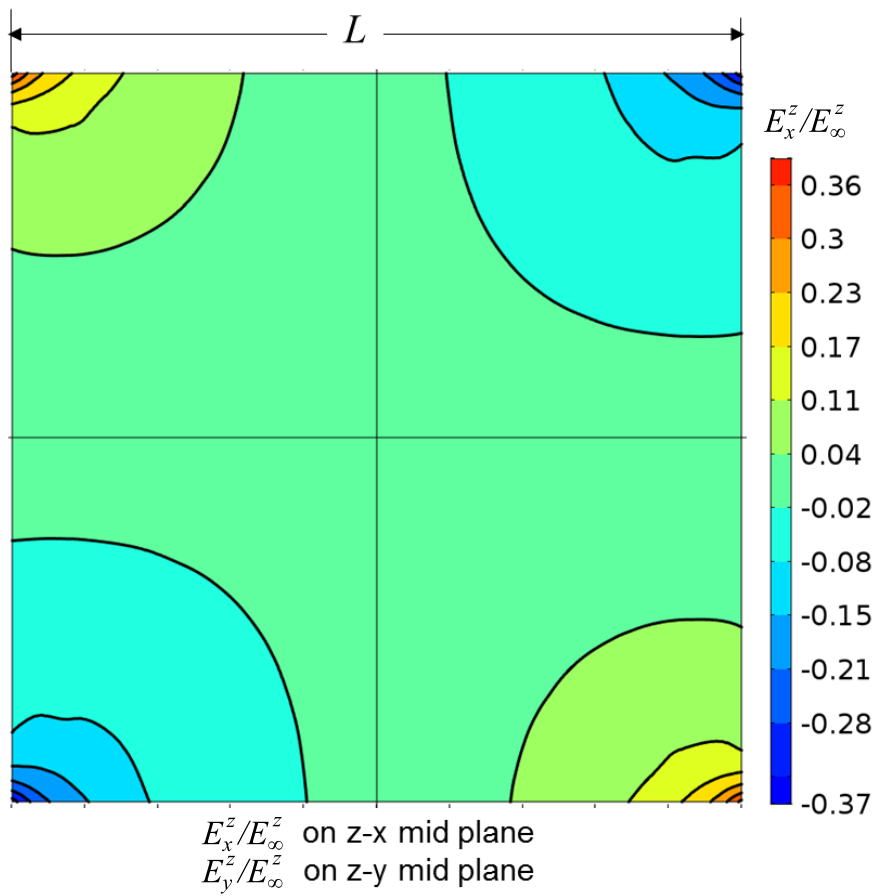

$E_x^z/E_\infty^z$ on z-x mid plane
$E_y^z/E_\infty^z$ on z-y mid plane

**Extended Data Figure 9 | Contour plot of normalized electric-field magnitude across a cube-shaped nanocrystal.** Contour plot of $E_x^z/E_\infty^z$ on the *xz* mid-plane of the cube. The ratio of the dielectric constant inside the nanocrystal ($\varepsilon_{in}$) to the surrounding medium ($\varepsilon_{out}$) was set to 9 (see Supplementary Section 3.B). The $E_y^z/E_\infty^z$ distribution on the *yz* mid-plane is identical. The *z* direction is vertical. The perturbations near the corners of the plots are artifacts of the interpolation resolution utilized by the software used to construct them.

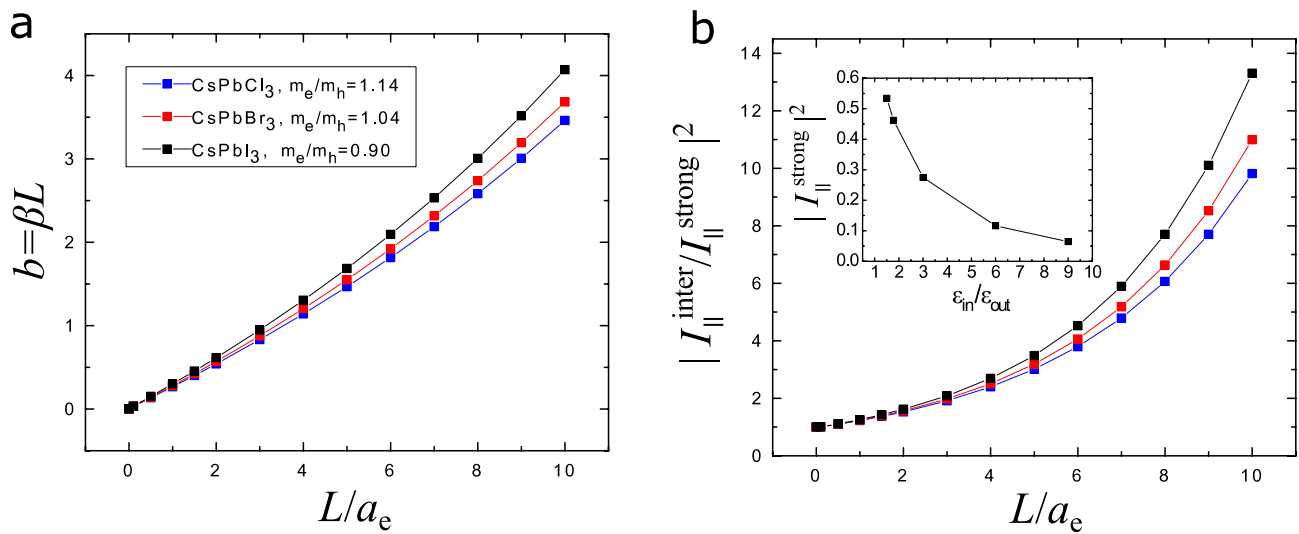

**Extended Data Figure 10 | Calculations related to the determination of the exciton radiative lifetime in cube-shaped nanocrystals within the intermediate-confinement regime. a**, Dimensionless electron–hole correlation constant, $b=\beta L$, and **b**, the square modulus of the ratio of $I_{\parallel}$ for intermediate and strong confinement as a function of the size of the nanocrystal relative to the electron Bohr radius, $L/a_e$, for the three materials studied. The inset in **b** shows the square modulus of $I_{\parallel}$ in the strong-confinement regime for several different dielectric constants, $\varepsilon_{in}/\varepsilon_{out}$. See Supplementary Section 3.D for details.

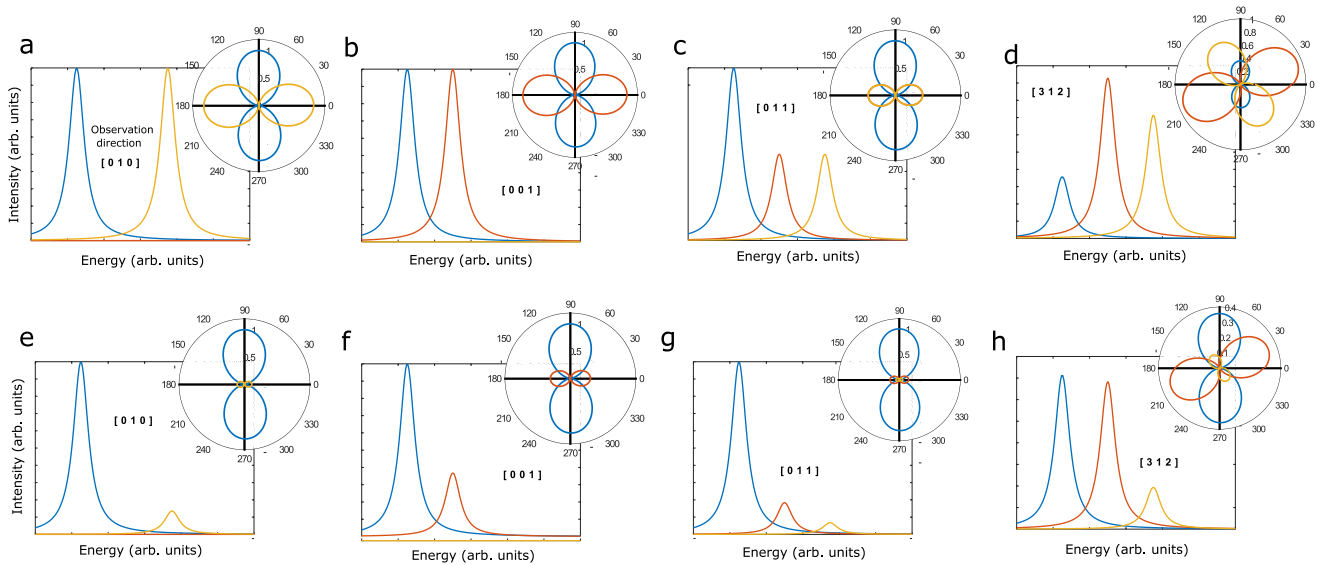

**Extended Data Figure 11 | Predicted exciton spectra and polarization properties for individual perovskite nanocrystals.** The plots show the expected exciton fine structure in photoluminescence spectra from three orthogonal dipoles of the lowest energy exciton. The dipoles are oriented along the orthorhombic symmetry axes. Each plot includes an inset with the emission probability for the dipoles as a function of the polarization angle. **a-d**, Expected fine structure for observation in the [010], [001], [011], and [312] directions with respect to the orthorhombic symmetry axes. The temperature effect on the population of the sublevels is not considered (*i.e.*, the populations of the sublevels are assumed to be equal). **e-h**, Expected fine structure for observation in the [010], [001], [011], and [312] directions with respect to the orthorhombic symmetry axes. The temperature effect on the population of the sublevels is considered. The temperature is assumed to be comparable to the fine-structure splitting, *i.e.* $k_bT \approx \Delta_1 = \Delta_2$, where $k_b$ is the Boltzmann constant and $T$ is temperature. See Supplementary Section 4 for further details.

# Supplementary information for:
# Bright triplet excitons in lead halide perovskites

Michael A. Becker, Roman Vaxenburg, Georgian Nedelcu, Peter C. Sercel, Andrew Shabaev, Michael J. Mehl, John G. Michopoulos, Samuel G. Lambrakos, Noam Bernstein, John L. Lyons, Thilo Stöferle, Rainer F. Mahrt, Maksym V. Kovalenko, David J. Norris, Gabriele Rainò, and Alexander L. Efros

## CONTENTS

## S1. EXCITON FINE STRUCTURE IN THE EFFECTIVE-MASS MODEL

### A. Four-band model

To calculate optical transition energies, exciton fine structure, polarization properties, as well as exciton and trion radiative lifetimes for perovskite nanocrystals, we need a multiband effective-mass Hamiltonian that describes the carrier energies in the vicinity of the bandgap edge in bulk perovskite semiconductors. Effective energy band parameters for this Hamiltonian can be extracted from the predicted electron and hole dispersion obtained via first-principle descriptions of the bulk energy structure for these semiconductors. Then, all the critical characteristics of perovskite nanocrystals can be calculated within this multiband effective-mass Hamiltonian. The goal of this and the next subsection is to develop this effective-mass Hamiltonian.

The R-point of the Brillouin zone is isomorphic to the $\Gamma$-point in cubic semiconductors [1]. As a result, the dispersion of electrons and holes at the R-point is described by the familiar 8x8 $\boldsymbol{k} \cdot \boldsymbol{p}$ Hamiltonian matrix that characterizes the band edge of direct-gap cubic semiconductors at the $\Gamma$-point. In the perovskites studied here, due to large spin–orbit coupling, a good description of the electron and hole dispersion is obtained by extracting the 4x4 part related to the $\Gamma_6^-$ and $\Gamma_6^+$ bands of the conduction and valence bands [2, 3]. Using the same standard semiconductor notation [4, 5] introduced in the main text, the Bloch wavefunctions of the corresponding band-edge states can be written as:

$$| \Uparrow \rangle_e = -\frac{1}{\sqrt{3}} \left[ (|X\rangle + i|Y\rangle)| \downarrow \rangle + |Z\rangle| \uparrow \rangle \right]$$

$$| \Downarrow \rangle_e = \frac{1}{\sqrt{3}} \left[ |Z\rangle| \downarrow \rangle - (|X\rangle - i|Y\rangle)| \uparrow \rangle \right]$$

$$| \uparrow \rangle_h = |S\rangle| \uparrow \rangle$$

$$| \downarrow \rangle_h = |S\rangle| \downarrow \rangle \ . \tag{S1}$$

Note the different phases of the basis functions in comparison with ref. 2. The Hamiltonian in our basis is

$$\hat{H} = \begin{pmatrix} E_c + \gamma_e \frac{p^2}{2m_0} & 0 & -\frac{1}{m_0} \frac{iP}{\sqrt{3}} p_z & -\frac{1}{m_0} \frac{iP}{\sqrt{3}} p_- \\ 0 & E_c + \gamma_e \frac{p^2}{2m_0} & -\frac{1}{m_0} \frac{iP}{\sqrt{3}} p_+ & \frac{1}{m_0} \frac{iP}{\sqrt{3}} p_z \\ \frac{1}{m_0} \frac{iP}{\sqrt{3}} p_z & \frac{1}{m_0} \frac{iP}{\sqrt{3}} p_- & E_v - \gamma_h \frac{p^2}{2m_0} & 0 \\ \frac{1}{m_0} \frac{iP}{\sqrt{3}} p_+ & -\frac{1}{m_0} \frac{iP}{\sqrt{3}} p_z & 0 & E_v - \gamma_h \frac{p^2}{2m_0} \end{pmatrix}, \tag{S2}$$

where $\hat{\boldsymbol{p}}$ is the momentum operator, $P = -i\langle S|p_z|Z\rangle$, $p_\pm = p_x \pm ip_y$, $p^2 = p_x^2 + p_y^2 + p_z^2$, $E_{c,v}$ are the band-edge energies, and $\gamma_{e,h}$ are the remote-band contributions to the electron and hole effective masses. Note that in the perovskites considered the band structure is reversed compared to many typical semiconductors in the sense that the valence band (instead of the conduction band) is $s$-like. The energy gap, $E_g = E_c - E_v$, is connected with the energy gap $E_g'$ and the spin–orbit splitting $\Delta$ of the standard 8-band model as $E_g = E_g' - |\Delta|$, because $\Delta$ in these perovskites is negative.

The energy spectrum of the carriers is isotropic at the R-point of the Brillouin zone and can be easily found by taking $\boldsymbol{p}$ along the $z$ axis. In this case, the $4 \times 4$ matrix is composed of two identical $2 \times 2$ blocks decoupled from each other, determined as,

$$\begin{pmatrix} E_g - \mathcal{E} + \dfrac{\gamma_e}{2m_0}p^2 & \dfrac{i}{\sqrt{3}}\dfrac{P}{m_0}p \\ -\dfrac{i}{\sqrt{3}}\dfrac{P}{m_0}p & -\mathcal{E} - \dfrac{\gamma_h}{2m_0}p^2 \end{pmatrix}. \tag{S3}$$

The usual procedures lead to the dispersion relation

$$\mathcal{E}_\pm = \frac{1}{2}\left[E_g + (\gamma_e - \gamma_h)\frac{p^2}{2m_0}\right] \pm \sqrt{\frac{1}{4}\left[E_g + (\gamma_e + \gamma_h)\frac{p^2}{2m_0}\right]^2 + E_p\frac{p^2}{6m_0}}, \tag{S4}$$

where we have used the Kane energy $E_p = 2P^2/m_0$.

## B. Estimating the energy-band parameters of the four-band model

To describe the energy spectra and radiative lifetimes in nanocrystals we need the parameters $E_p$ and $\gamma_{e,h}$ of the four-band model for the cubic perovskites, Eq.(S4). We determine these parameters by fitting the first-principles calculations presented in Fig. 1b in the main text and Extended Data Fig. 1. In particular, $E_p$ and $\gamma_{e,h}$ are connected to the effective masses of the electrons and holes, $m_e$ and $m_h$, at their respective band edges by the following relationships:

$$\frac{m_0}{m_e} = \gamma_e + \frac{E_p}{3E_g} \ , \ \frac{m_0}{m_h} = \gamma_h + \frac{E_p}{3E_g}. \tag{S5}$$

These expressions are derived from the parabolic approximation to Eq.(S4) applied for small $p$ and using $E_p \gg E_g$. To extract $E_p$, we take the asymptotic limit of Eq.(S4) at large $p$, such that $p^2 E_p \gg 6m_0 E_g^2$. Assuming that $E_p$ is sufficiently large that $p^2 \ll m_0 E_p/(\gamma_e \pm \gamma_h)$, which is satisfied for a very wide range of energies in the conduction and valence bands, we obtain,

$$\mathcal{E} \approx \frac{1}{2}E_g \pm \sqrt{\frac{E_p}{6m_0}}p, \ \left(p^2 \gg \frac{6m_0 E_g^2}{E_p}\right). \tag{S6}$$

|  | CsPbCl$_3$ | CsPbBr$_3$ | CsPbI$_3$ |
|---|---|---|---|
| $E_g$ exp (eV) | 3.04 [6] | 2.36 [6] | 1.67 [7] |
| $E_p$ (eV) | 40.1 | 39.9 | 41.6 |
| $\gamma_e$ | 0.77 | 1.85 | 3.27 |
| $\gamma_h$ | 1.51 | 2.21 | 2.58 |
| $m_e/m_0$ | 0.194 | 0.134 | 0.086 |
| $m_h/m_0$ | 0.170 | 0.128 | 0.095 |
| $\epsilon_{in}$ exp. | 4.5 | 4.8 | 5.0 |

TABLE S1. Parameters of the four-band model, describing the dispersion of the conduction and valence bands in the vicinity of the R-point of the Brillouin zone, and the high-frequency dielectric constants of the CsPbX$_3$ (X=Cl, Br, and I) perovskites used in calculations of the exciton lifetimes. The band parameters were extracted from first-principles calculations of the perovskite band structures. The high-frequency dielectric constants were obtained from an analysis of the exciton binding energy and the refractive index.

Extended Data Fig. 6 shows the slope $\sqrt{E_p/6m_0}$ according to Eq.(S6), which is calculated as the energy difference $\mathcal{E} - E_g/2$ divided by the corresponding difference in momentum, $p = \hbar\Delta k$, where $\Delta k$ is the wave number shown in nm$^{-1}$. The results of this fitting procedure are summarized in the Table S1 along with the energy gaps $E_g$ of the bulk perovskites CsPbX$_3$ taken from experimental data.

The last parameter needed to analyze the exciton radiative lifetimes is the high-frequency dielectric constant for each material. Using the effective masses summarized in Table S1 we can calculate the high-frequency dielectric constant from the exciton Rydberg when it is known. Taking the measured Rydberg for CsPbCl$_3$ and CsPbBr$_3$, 60 meV and 34 meV, respectively [6], we find $\epsilon_{in} = 4.5$ for CsPbCl$_3$ and $\epsilon_{in} = 4.8$ in CsPbBr$_3$. The Rydberg has not been measured for CsPbI$_3$. Noting that the dielectric constants determined for CsPbCl$_3$ and CsPbBr$_3$ are very close to those measured in methylammonium (MA) lead halide perovskites [8] ($\epsilon_{in} = 4.0$ in MAPbCl$_3$ and $\epsilon_{in} = 4.7$ in MAPbBr$_3$), we have taken the dielectric constant of MAPbI$_3$, $\epsilon_{in} = 5.0$ [8], for CsPbI$_3$. The dielectric constants and the energy-band parameters from Table S1 are used in the calculations of the radiative lifetimes.

## C. Exciton fine structure in nanocrystals with cubic lattice structure

The total wavefunction of the electron and hole states in nanocrystals can be found using the parabolic-band effective-mass approximation [9, 10]. They generally can be written as a product of the Bloch functions defined in Eq.(S1) and envelope functions, *i.e.* products of the form: $\Psi^e_{\Uparrow,\Downarrow}(\boldsymbol{r}_e) = F_e(\boldsymbol{r}_e)| \Uparrow, \Downarrow\rangle_e$ and $\Psi^h_{\uparrow,\downarrow}(\boldsymbol{r}_h) = F_h(\boldsymbol{r}_h)| \uparrow, \downarrow\rangle_h$ for electron and hole states, respectively, where $F_{e,h}$ are the electron or hole envelope functions.

The total exciton wavefunctions in nanocrystals are the product of the Bloch functions defined in Eqs.(2) and (3) in the main text and the exciton envelope function $v(\boldsymbol{r}_e, \boldsymbol{r}_h)$, which describes spatial motion of the exciton confined in the nanocrystal. The resulting wavefunctions of the exciton $\Psi^{ex}_{J,J_z}$ with momentum $J$ and momentum projection $J_z$ have the following form:

$$\Psi^{ex}_{0,0} = \frac{1}{\sqrt{2}} \left(| \Downarrow\rangle_e| \uparrow\rangle_h - | \Uparrow\rangle_e| \downarrow\rangle_h\right) v(\boldsymbol{r}_e, \boldsymbol{r}_h) \ , \ \ \Psi^{ex}_{1,+1} = | \Uparrow\rangle_e| \uparrow\rangle_h v(\boldsymbol{r}_e, \boldsymbol{r}_h) \ ,$$

$$\Psi^{ex}_{1,0} = \frac{1}{\sqrt{2}} \left(| \Downarrow\rangle_e| \uparrow\rangle_h + | \Uparrow\rangle_e| \downarrow\rangle_h\right) v(\boldsymbol{r}_e, \boldsymbol{r}_h) \ , \ \ \Psi^{ex}_{1,-1} = | \Downarrow\rangle_e| \downarrow\rangle_h v(\boldsymbol{r}_e, \boldsymbol{r}_h) \ . \quad (S7)$$

The electron–hole exchange interaction [1], $H_{\text{exch}} = -\alpha_{exc} \, \Omega_0 (\boldsymbol{\sigma}^e \cdot \boldsymbol{\sigma}^h) \delta(\boldsymbol{r}_e - \boldsymbol{r}_h)$, where $\boldsymbol{\sigma}^{e,h}$ are the electron and hole Pauli operators, $\alpha_{exc}$ is the exchange constant, and $\Omega_0$ is the volume of the unit cell, conserves the two-particle angular momentum, $\boldsymbol{J} = \frac{1}{2}(\boldsymbol{\sigma}^e + \boldsymbol{\sigma}^h)$. This exchange interaction splits the fourfold degenerate exciton ground state into an optically passive singlet ($J = 0$) and a threefold degenerate optically active triplet state ($J = 1$ with three momentum projections $J_z = \pm 1, 0$).

The singlet–triplet splitting of the exciton levels can be shown to be equal to $4\eta$, where $\eta = \alpha_{exc}\Theta$, with $\Theta = \Omega_0 \int d^3 r v^2(r, r)$. It is known that the splitting is enhanced by spatial confinement [11], which is included via the parameter $\Theta$. In the strong-confinement regime: $\Theta \sim \Omega_0/V$ and is inversely proportional to the nanocrystal volume, $V$. In the bulk and in the weak-confinement regime: $\Theta = \Omega_0/\pi a_B^3$.

It is easy to demonstrate that the singlet level $|\Psi_{0,0}\rangle$ is optically passive. This is because the transition-dipole matrix element taken between this state and the vacuum state $|0\rangle = \delta(\mathbf{r}_e - \mathbf{r}_h)$ is zero: $\langle 0|\hat{\mathbf{p}}|\Psi_{0,0}\rangle = 0$. In the optical matrix element, $\hat{\mathbf{p}}$ acts only on the conduction-band Bloch functions. Thus, the exciton wavefunction $|\Psi_{0,0}\rangle$ from Eq.(2) in the main text should be transformed to the electron–electron representation. In this case, using the time-reversal operator $\hat{K}$ for transformation of the hole wavefunction to the electron

form, one can show that

$$
\begin{aligned}
\langle 0|\hat{\mathbf{p}}|\Psi_{0,0}\rangle &= \int d^3r \frac{1}{\sqrt{2}} \left( (\hat{K}|S\rangle|\uparrow\rangle) \hat{\mathbf{p}} \frac{1}{\sqrt{3}} \left[ |Z\rangle|\downarrow\rangle - (|X\rangle - i|Y\rangle)|\uparrow\rangle \right] \right. \\
&\quad \left. - (\hat{K}|S\rangle|\downarrow\rangle) \hat{\mathbf{p}} \left( -\frac{1}{\sqrt{3}} \left[ (|X\rangle + i|Y\rangle)|\downarrow\rangle + |Z\rangle|\uparrow\rangle \right] \right) \right) \\
&= \frac{1}{\sqrt{6}} \int d^3r \left( \langle S|\langle \downarrow| \hat{\mathbf{p}} \left[ |Z\rangle|\downarrow\rangle - (|X\rangle - i|Y\rangle)|\uparrow\rangle \right] \right. \\
&\quad \left. + \langle S|\langle \uparrow| \hat{\mathbf{p}} \left( - \left[ (|X\rangle + i|Y\rangle)|\downarrow\rangle + |Z\rangle|\uparrow\rangle \right] \right) \right) \\
&= \frac{1}{\sqrt{6}} \int d^3r \left( \langle S|\hat{\mathbf{p}}|Z\rangle - \langle S|\hat{\mathbf{p}}|Z\rangle \right) = 0 \ .
\end{aligned}
\tag{S8}
$$

Here we used the following properties of the time-reversal operator $\hat{K}$: $\hat{K}|\uparrow\rangle = |\downarrow\rangle$ and $\hat{K}|\downarrow\rangle = -|\uparrow\rangle$. Similar calculations show that all three triplet states are optically active.

### D. The order of the singlet and triplet excitons in perovskite nanocrystals

The sign of the exchange-interaction constant $\alpha_{exc}$ affects the level order of the singlet and triplet exciton states. In the absence of spin–orbit coupling, both $\alpha_{exc}$ and $\eta$ are always positive, resulting in a optically passive spin-triplet exciton ground state. This is the case for organic semiconductors. When strong spin–orbit coupling exists in only one band (for which the corresponding band-edge Bloch functions are described by Eq. 1 of the main text and above), the parameters $\alpha_{exc}$ and $\eta$ are negative leading to an *optically passive singlet exciton ground state.* Ignoring the Rashba effect for the moment (see the next subsection, Section S1.E), this optically passive singlet would be the expected exciton ground state for perovskites and perovskite nanocrystals [12–16]. The splitting in this case was intensively analyzed theoretically [17–19] in connection with CuCl, for which the conduction and valence band edges have symmetry $\Gamma_6^+$ and $\Gamma_6^-$, respectively. The triplet–singlet splitting, $4\eta$ can be expressed in terms of the Bloch functions of the conduction and valence bands [18]: $4\eta = (2/3)(\Theta/\Omega_0^2) \int d^3r_1 d^3r_2 S^*(\boldsymbol{r_1}) X^*(\boldsymbol{r_2}) V(\boldsymbol{r_1} - \boldsymbol{r_2}) S(\boldsymbol{r_2}) X(\boldsymbol{r_1})$, where the Bloch functions are normalized to the unit-cell volume, $\Omega_0$, the integrals are taken over one unit cell, and $V(\boldsymbol{r_1} - \boldsymbol{r_2}) = e^2/(\epsilon_{in}|\boldsymbol{r_1} - \boldsymbol{r_2}|)$ is the Coulomb potential between two electrons. This exchange integral is *always* positive and the optically active triplet *always* has higher energy.

To quantify the exchange splitting for perovskite nanocrystals (still ignoring the Rashba effect), we conducted first-principles calculations of the band-edge Bloch functions and calculated the exchange constant $\alpha_{exc}$ for all three CsPbX$_3$ halide perovskites. The results of these calculations, using Heyd-Scuseria-Ernzerhof HSE06 hybrid functionals, which mix exact Hartree-Fock exchange with conventional DFT (see Methods), are shown for CsPbX$_3$ (X=Cl, Br, I) in Table S2. The calculated short-range exchange splitting of the singlet–triplet exciton is shown in the third column of the table.

TABLE S2. Calculated exchange constant $\alpha_{exc}$, $\Theta$, and singlet–triplet splitting $4\eta = 4\alpha_{exc}\Theta$ in CsPbX$_3$ (X=Cl, Br, and I).

| Perovskite | $4\alpha_{exc}$(meV) | $\Theta$ | singlet–triplet splitting, $4\eta$ (meV) |
|:---:|:---:|:---:|:---:|
| CsPbCl$_3$ | 354 | 0.00313 | 1.11 |
| CsPbBr$_3$ | 267 | 0.00110 | 0.29 |
| CsPbI$_3$ | 204 | 0.00038 | 0.078 |

Our experimentally studied CsPbX$_3$ nanocrystals are known to exhibit an orthorhombic lattice distortion [20]. The reduction of the nanocrystal symmetry generally splits the threefold degenerate triplet states into three exciton sublevels. To find these splittings in CsPbBr$_3$ nanocrystals, we used $G_0W_0$ first-principle calculations (see Methods). Our calculations predict an expected orthorhombic splitting for the triplet with $\Delta_1 = 1.9 * 0.004388 = 0.008$ meV and $\Delta_2 = 3.9 * 0.004388 = 0.017$ meV (see inset to Fig. 3i in the main text for definitions of $\Delta_1$ and $\Delta_2$). These splittings are hundreds of times smaller than the splittings measured experimentally ($\sim 1$ meV) in our perovskite nanocrystals (see Fig. 3). This suggests that the orthorhombic distortion is not responsible for the observed splittings.

### E. Effect of Rashba terms on the exciton fine structure

We now consider the influence of the Rashba effect on the observed exciton fine structure. This effect can arise due to inversion-symmetry breaking in CsPbBr$_3$, for example due to the instability of Cs$^+$ ions in the lattice [21]. Instabilities in the ion positions can result in lattice polarization, which creates Rashba terms in the Hamiltonians describing the electrons and holes.

The Rashba effect for electrons and holes in a nanocrystal made from a cubic crystal lattice can be described as [22]:

$$\hat{H}_R^c = \alpha_{R,c}[(\sigma_x p_y - \sigma_y p_x)n_z + (\sigma_z p_x - \sigma_x p_z)n_y + (\sigma_y p_z - \sigma_z p_y)n_x] \ , \tag{S9}$$

where $\alpha_{R,c}$ represents either the conduction- or valence-band Rashba coefficient for nanocrystals with cubic lattice structure ($\alpha_{R,c}^e$ and $\alpha_{R,c}^h$, respectively), and $\sigma_i$ are the projections of the Pauli operators for the electron total momentum operator for $J = 1/2$ and for the hole spin $s = 1/2$, ($\sigma_i^e$ and $\sigma_i^h$ respectively). In Eq. (S9), $n_{x,y,z}$ are the projections on the cubic axes of a unit vector $\boldsymbol{n}$ defining the direction of the symmetry breaking (see, for example, the inset in Fig. 1c of the main text).

From Eq. (S9) we can write the most general Rashba Hamiltonian for nanocrystals with orthorhombic symmetry:

$$\hat{H}_R^o = \alpha_{R,xy}^z \sigma_x p_y n_z - \alpha_{R,yx}^z \sigma_y p_x n_z + \alpha_{R,zx}^y \sigma_z p_x n_y - \alpha_{R,xz}^y \sigma_x p_z n_y$$
$$+ \ \alpha_{R,yz}^x \sigma_y p_z n_x - \alpha_{R,zy}^x \sigma_z p_y n_x \ . \tag{S10}$$

As one can see from Eq. (S10), the Rashba effect for both electrons and holes is fully described with six independent parameters: $\alpha_{R,xy}^{z,e;h}$, $\alpha_{R,yx}^{z,e;h}$, $\alpha_{R,zx}^{y,e;h}$, $\alpha_{R,xz}^{y,e;h}$, $\alpha_{R,yz}^{x,e;h}$, and $\alpha_{R,zy}^{x,e;h}$, respectively, which reflect the material properties and symmetry of the nanocrystal, while again the projections $n_x, n_y, n_z$ of the unit vector $\boldsymbol{n}$ define the Rashba symmetry-breaking direction. For calculations it is convenient to re-write the Rashba Hamiltonian in Eq. (S10) acting on the exciton as a sum of the three terms $\hat{H}_R^o = \hat{H}_x n_x + \hat{H}_y n_y + \hat{H}_z n_z$:

$$\hat{H}_z = \alpha_{R,xy}^{z,e} \sigma_x^e p_y^e - \alpha_{R,yx}^{z,e} \sigma_y^e p_x^e + \alpha_{R,xy}^{z,h} \sigma_x^h p_y^h - \alpha_{R,yx}^{z,h} \sigma_y^h p_x^h \ ,$$
$$\hat{H}_y = \alpha_{R,zx}^{y,e} \sigma_z^e p_x^e - \alpha_{R,xz}^{y,e} \sigma_x^e p_z^e + \alpha_{R,zx}^{y,h} \sigma_z^h p_x^h - \alpha_{R,xz}^{y,h} \sigma_x^h p_z^h \ ,$$
$$\hat{H}_x = \alpha_{R,yz}^{x,e} \sigma_y^e p_z^e - \alpha_{R,zy}^{x,e} \sigma_z^e p_y^e + \alpha_{R,yz}^{x,h} \sigma_y^h p_z^h - \alpha_{R,zy}^{x,h} \sigma_z^h p_y^h \ . \tag{S11}$$

The wavefunction of the exciton ground state in cube-shaped nanocrystals in the weak confinement regime (which we use to approximate our experimental samples), can be written as:

$$\Psi_{gr}^{J,J_z}(\boldsymbol{R}, \boldsymbol{r}) = \psi_{100}(\boldsymbol{r})\Psi_0(\boldsymbol{R})U_{\boldsymbol{J},J_z} \tag{S12}$$

Here $\psi_{100}(\boldsymbol{r})$ is the hydrogen-like function that describes the relative motion of the exciton ground state with $\boldsymbol{r} = \boldsymbol{r}_h - \boldsymbol{r}_e$. For the ground state, the wavefunction of the exciton relative motion can be written:

$$\psi_{100}(\mathbf{r}) = 2 \left( \frac{1}{a_B} \right)^{3/2} e^{-r/a_B} \ Y_{0,0}, \tag{S13}$$

where $a_B = \epsilon_{\text{in}} \hbar^2 / e^2 / \mu$ is the exciton Bohr radius with $\mu = [1/[1/m_e + 1/m_h]$ as the reduced exciton effective mass, and $Y_{0,0}$ is the spherical harmonic with $l = 0$. $\Psi_0(\boldsymbol{R})$ describes the wavefunction of the exciton center-of-mass motion, with $\boldsymbol{R} = [m_e \boldsymbol{r}_e + m_h \boldsymbol{r}_h]/M$, where $m_e$ and $m_h$ are the electron and hole effective masses, and $M = m_e + m_h$. For the exciton ground state, the wavefunction of the exciton center-of-mass motion can be written:

$$\Psi_0(\boldsymbol{R}) = \sqrt{\frac{8}{L_x L_y L_z}} \cos(\pi X / L_x) \cos(\pi Y / L_y) \cos(\pi Z / L_z), \tag{S14}$$

where $L_x$, $L_y$, and $L_z$ are the edge lengths of the cube-shaped nanocrystal. Finally $U_{\boldsymbol{J}, J_z}$ in Eq.(S12) is the spin part of the exciton function, which for $J = 0$ and $J = 1$ (singlet and triplet states, respectively) can be written as:

$$|0, 0\rangle = \frac{1}{\sqrt{2}} (|\Uparrow\rangle| \downarrow\rangle - |\Downarrow\rangle| \uparrow\rangle) \tag{S15}$$

$$|1, +1\rangle = |\Uparrow\rangle| \uparrow\rangle, \qquad |1, 0\rangle = \frac{1}{\sqrt{2}} (|\Uparrow\rangle| \downarrow\rangle + |\Downarrow\rangle| \uparrow\rangle)), \qquad |1, -1\rangle = |\Downarrow\rangle| \downarrow\rangle . \tag{S16}$$

Corrections to the exciton ground state from the Rashba terms in Eqs. (S11) vanish in first-order perturbation theory. In second-order perturbation theory, however, we find corrections that describe coupling among the spin sublevels of the exciton. The resulting coupling matrix contains spin–spin coupling terms and is similar in that respect to an effective exchange Hamiltonian. In second-order perturbation theory this matrix can be written [23]:

$$M_{J, J_z}^{J', J_z'} = \sum_{m; J'', J_z''} \frac{\langle \Psi_{gr}^{J, J_z} | \hat{H}_R | \Psi_m^{J'', J_z''} \rangle \langle \Psi_m^{J'', J_z''} | \hat{H}_R | \Psi_{gr}^{J', J_z'} \rangle}{E_{gr} - E_m}, \tag{S17}$$

where the sum goes over all intermediate spatial states $m$ and all spin states $J''$ and $J_z''$. However, significant simplifications arise because the energies of the intermediate states in this expression are independent of spin. One can sum over all intermediate $J'', J_z''$ states, resulting in a 4x4 coupling matrix.

To estimate the matrix in Eq. (S17) we take into account just the first few excited states of the exciton center-of-mass and relative motions. In the later case, the wavefunction can be written as:

$$\Psi_{1m}^{r; J, J_z}(\boldsymbol{R}, \boldsymbol{r}) = \psi_{21m}(\boldsymbol{r}) \Psi_0(\boldsymbol{R}) U_{\boldsymbol{J}, J_z}, \tag{S18}$$

where $\psi_{21m}(\boldsymbol{r})$ is the hydrogen-like wavefunction of the 1P exciton level with angular momentum $l = 1$ and momentum projections $m = 0, \pm 1$. These wavefunctions can be written

as:

$$\psi_{21m}(\boldsymbol{r}) = \frac{1}{2\sqrt{6}} \left(\frac{1}{a_B}\right)^{3/2} \frac{r}{a_B} e^{-r/2a_B} \, Y_{1,m}(\theta,\phi) \;, \tag{S19}$$

where $Y_{1,m}$ are the spherical harmonics with $l=1$ [23]. The energy distance for the 1P level is $0.75e^4\mu/\hbar^2\epsilon_{in}^2$.

For the first three excited levels connected with the exciton center-of-mass motion we can write:

$$\Psi^{R,J,J_z}_{x,y,z}(\boldsymbol{R},\boldsymbol{r}) = \psi_{100}(\boldsymbol{r})\Psi_{x,y,z}(\boldsymbol{R})U_{\boldsymbol{J},J_z} \;, \tag{S20}$$

where the excited wavefunction of the exciton center of mass motion $\Psi_{x,y,z}$ can be written:

$$\Psi_x(\boldsymbol{R}) = \sqrt{\frac{8}{L_xL_yL_z}} \sin(2\pi X/L_x)\cos(\pi Y/L_y)\cos(\pi Z/L_z),$$

$$\Psi_y(\boldsymbol{R}) = \sqrt{\frac{8}{L_xL_yL_z}} \cos(\pi X/L_x)\sin(2\pi Y/L_y)\cos(\pi Z/L_z),$$

$$\Psi_z(\boldsymbol{R}) = \sqrt{\frac{8}{L_xL_yL_z}} \cos(\pi X/L_x)\cos(\pi Y/L_y)\sin(2\pi Z/L_z). \tag{S21}$$

The energy distances between the ground and excited exciton states are $3\hbar^2\pi^2/2ML_x^2$, $3\hbar^2\pi^2/2ML_y^2$, and $3\hbar^2\pi^2/2ML_z^2$, respectively.

Let us now calculate the effective electron–hole spin-coupling Hamiltonian created by the Rashba term. Substituting $\hat{H}^o_R$ into Eq. (S17) we obtain:

$$M^{J',J'_z}_{J,J_z} =$$
$$- \frac{2m_e}{M}(A^c_Rm_e + A^r_Rm_h)\left[(\tilde\alpha^{y,e}_{R,zx})^2 + (\tilde\alpha^{z,e}_{R,yx})^2 + (\tilde\alpha^{z,e}_{R,xy})^2 + (\tilde\alpha^{x,e}_{R,zy})^2 + (\tilde\alpha^{x,e}_{R,yz})^2 + (\tilde\alpha^{y,e}_{R,xz})^2\right]$$
$$- \frac{2m_h}{M}(A^c_Rm_h + A^r_Rm_e)\left[(\tilde\alpha^{y,h}_{R,zx})^2 + (\tilde\alpha^{z,h}_{R,yx})^2 + (\tilde\alpha^{z,h}_{R,xy})^2 + (\tilde\alpha^{x,h}_{R,zy})^2 + (\tilde\alpha^{x,h}_{R,yz})^2 + (\tilde\alpha^{y,h}_{R,xz})^2\right]$$
$$- \frac{2m_hm_e}{M}(A^c_R - A^r_R)$$
$$\times \left[\left(\tilde\alpha^{z,e}_{R,xy}\tilde\alpha^{z,h}_{R,xy} + \tilde\alpha^{y,e}_{R,xz}\tilde\alpha^{y,h}_{R,xz}\right)\sigma^e_x\sigma^h_x + \left(\tilde\alpha^{x,e}_{R,yz}\tilde\alpha^{x,h}_{R,yz} + \tilde\alpha^{z,e}_{R,yx}\tilde\alpha^{z,h}_{R,yx}\right)\sigma^e_y\sigma^h_y\right.$$
$$+ \left(\tilde\alpha^{y,e}_{R,zx}\tilde\alpha^{y,h}_{R,zx} + \tilde\alpha^{x,e}_{R,zy}\tilde\alpha^{x,h}_{R,zy}\right)\sigma^e_z\sigma^h_z - \tilde\alpha^{y,e}_{R,xz}\tilde\alpha^{x,h}_{R,yz}\sigma^e_x\sigma^h_y - \tilde\alpha^{x,e}_{R,yz}\tilde\alpha^{y,h}_{R,xz}\sigma^e_y\sigma^h_x - \tilde\alpha^{z,e}_{R,xy}\tilde\alpha^{x,h}_{R,zy}\sigma^e_x\sigma^h_z$$
$$\left. - \tilde\alpha^{x,e}_{R,zy}\tilde\alpha^{z,h}_{R,xy}\sigma^e_z\sigma^h_x - \tilde\alpha^{y,e}_{R,zx}\tilde\alpha^{z,h}_{R,yx}\sigma^e_z\sigma^h_y - \tilde\alpha^{z,e}_{R,yx}\tilde\alpha^{y,h}_{R,zx}\sigma^e_y\sigma^h_z\right], \tag{S22}$$

where, $\tilde\alpha^{z,e;h}_{R,ij} = \alpha^{z,e;h}_{R,ij}n_z$, $\tilde\alpha^{y,e;h}_{R,ij} = \alpha^{y,e;h}_{R,ij}n_y$, $\tilde\alpha^{x,e;h}_{R,ij} = \alpha^{x,e;h}_{R,ij}n_x$, $A^c_R = 128/(27\pi^2)$, and $A^r_R = (64/81\sqrt{3})^2$. The terms proportional to $A^c_R$ and $A^r_R$ come from the intermediate states connected with the exciton center-of-mass motion and the relative motion of the electron and hole, respectively. The third term in Eq. (S22) consists of spin–spin coupling terms and has the same form as the effective spin-dependent electron–hole exchange Hamiltonian. Such terms determine the fine structure of the band-edge exciton. One can see that the contributions of the center-of-mass motion and the relative motion of the exciton have different signs and result in a different level order. However, because $A_R^c > A_R^r$ it is the center-of-mass motion that determines the level order of the exciton.

The fine structure of the exciton is thus defined by the following matrix:

$$\hat{H} = -(A_R^c - A_R^r)\frac{2m_e m_h}{M} \times$$

$$
\begin{array}{cccc}
|\Uparrow\uparrow\rangle & |\Uparrow\downarrow\rangle & |\Downarrow\uparrow\rangle & |\Downarrow\downarrow\rangle
\end{array}
$$

$$
\begin{pmatrix}
\tilde{\alpha}_{R,zy}^{x,e}\tilde{\alpha}_{R,zy}^{x,h} + \tilde{\alpha}_{R,zz}^{y,e}\tilde{\alpha}_{R,zz}^{y,h} & -\tilde{\alpha}_{R,zy}^{x,e}\tilde{\alpha}_{R,xy}^{z,h} + i\tilde{\alpha}_{R,zz}^{y,e}\tilde{\alpha}_{R,yx}^{z,h} & -\tilde{\alpha}_{R,zy}^{x,e}\tilde{\alpha}_{R,zy}^{x,h} + i\tilde{\alpha}_{R,yx}^{z,e}\tilde{\alpha}_{R,zz}^{y,h} & M_{14} \\
-\tilde{\alpha}_{R,xy}^{x,e}\tilde{\alpha}_{R,xy}^{z,h} - i\tilde{\alpha}_{R,zz}^{y,e}\tilde{\alpha}_{R,yx}^{z,h} & -\tilde{\alpha}_{R,zy}^{x,e}\tilde{\alpha}_{R,zy}^{x,h} - \tilde{\alpha}_{R,zz}^{y,e}\tilde{\alpha}_{R,zz}^{y,h} & M_{23} & \tilde{\alpha}_{R,xy}^{z,e}\tilde{\alpha}_{R,zy}^{x,h} - i\tilde{\alpha}_{R,zz}^{z,e}\tilde{\alpha}_{R,zz}^{y,h} \\
-\tilde{\alpha}_{R,xy}^{z,e}\tilde{\alpha}_{R,zy}^{x,h} - i\tilde{\alpha}_{R,yx}^{z,e}\tilde{\alpha}_{R,zz}^{y,h} & M_{32} & -\tilde{\alpha}_{R,zy}^{x,e}\tilde{\alpha}_{R,zy}^{x,h} - \tilde{\alpha}_{R,zz}^{y,e}\tilde{\alpha}_{R,zz}^{y,h} & \tilde{\alpha}_{R,zy}^{x,e}\tilde{\alpha}_{R,xy}^{z,h} - i\tilde{\alpha}_{R,zz}^{y,e}\tilde{\alpha}_{R,yx}^{z,h} \\
M_{41} & \tilde{\alpha}_{R,xy}^{z,e}\tilde{\alpha}_{R,zy}^{x,h} + i\tilde{\alpha}_{R,yx}^{z,e}\tilde{\alpha}_{R,zz}^{y,h} & \tilde{\alpha}_{R,zy}^{x,e}\tilde{\alpha}_{R,xy}^{z,h} + i\tilde{\alpha}_{R,zz}^{y,e}\tilde{\alpha}_{R,yx}^{z,h} & \tilde{\alpha}_{R,zy}^{x,e}\tilde{\alpha}_{R,zy}^{x,h} + \tilde{\alpha}_{R,zz}^{y,e}\tilde{\alpha}_{R,zz}^{y,h}
\end{pmatrix},
$$

$$(S23)$$

where:

$$
\begin{aligned}
M_{14} = M_{41}^* &= \tilde{\alpha}_{R,xy}^{z,e}\tilde{\alpha}_{R,xy}^{z,h} - \tilde{\alpha}_{R,yx}^{z,e}\tilde{\alpha}_{R,yx}^{z,h} + (\tilde{\alpha}_{R,xz}^{y,e} + i\tilde{\alpha}_{R,yz}^{x,e})(\tilde{\alpha}_{R,xz}^{y,h} + i\tilde{\alpha}_{R,yz}^{x,h}) \,, \\
M_{23} = M_{32}^* &= \tilde{\alpha}_{R,xy}^{z,e}\tilde{\alpha}_{R,xy}^{z,h} + \tilde{\alpha}_{R,yx}^{z,e}\tilde{\alpha}_{R,yx}^{z,h} + (\tilde{\alpha}_{R,xz}^{y,e} + i\tilde{\alpha}_{R,yz}^{x,e})(\tilde{\alpha}_{R,xz}^{y,h} - i\tilde{\alpha}_{R,yz}^{x,h}) \,.
\end{aligned}
$$

$$(S24)$$

The exciton fine level structure can be found analytically from Eq. (S23) in the case when the Rashba anisotropy axis $\boldsymbol{n}$ is aligned along one symmetry axis of the nanocrystal ( $n_x = n_y = 0$, and $|n_z| = 1$ ). In this case, the energy level structure and the level polarization can be written:

$$
\begin{array}{lll}
\text{Energy} & & \text{Polarization} \\
\epsilon_d = \alpha_{R,xy}^{z,e}\alpha_{R,xy}^{z,h} + \alpha_{R,yx}^{z,e}\alpha_{R,yx}^{z,h} & & \text{dark} = |\text{d}\rangle \\
\epsilon_x = \alpha_{R,xy}^{z,e}\alpha_{R,xy}^{z,h} - \alpha_{R,yx}^{z,e}\alpha_{R,yx}^{z,h} & & |x\rangle \\
\epsilon_y = -\alpha_{R,xy}^{z,e}\alpha_{R,xy}^{z,h} + \alpha_{R,yx}^{z,e}\alpha_{R,yx}^{z,h} & & |y\rangle \\
\epsilon_z = -\alpha_{R,xy}^{z,e}\alpha_{R,xy}^{z,h} - \alpha_{R,yx}^{z,e}\alpha_{R,yx}^{z,h} & & |z\rangle \,,
\end{array}
$$

$$(S25)$$

where the energy is in units of $(A_R^c - A_R^r)(2m_e m_h/M)$. In perovskite nanocrystals the Rashba coefficient for both electrons $\alpha_{R,jk}^{i,e}$ and holes $\alpha_{R,jk}^{i,h}$ are negative as estimated using expressions derived from third-order perturbation theory within an extended Kane model [24]. As a result, the upper exciton state is the dark exciton and the lowest of the three optically active states is polarized along the anisotropy direction $z$. Generally, the two intermediate levels are split and have $x$ and $y$ polarization.

However, in the case when $\alpha_{R,xy}^{z,e} = \alpha_{R,yx}^{z,e}$ and $\alpha_{R,xy}^{z,h} = \alpha_{R,yx}^{z,h}$ , which occurs in nanocrystals with two equivalent symmetry axes, the $x$ and $y$ lines become degenerate and create a circularly polarized doublet, with polarization $x \pm iy$. The eigenvalues from Eq. (S25) are reduced to $\epsilon_x = \epsilon_y = 0$ and $\epsilon_z = -\epsilon_d$. This level structure we believe was observed recently [16]. The Rashba splitting between the two bright excitons in that case is described as $2m_e m_h (A_R^c - A_R^r)\epsilon_d/M$.

Another analytical expression for the exciton fine structure can be found for the case when $n_z = 0$ and $n_x, n_y \neq 0$. Diagonalization of the matrix described by Eq. (S23) gives energy levels, again in units of $2m_e m_h (A_R^c - A_R^r)\epsilon_d/M$, which can be written as:

$$\epsilon_1 = \tilde{\alpha}_{R,zy}^{x,e} \tilde{\alpha}_{R,zy}^{x,h} + \tilde{\alpha}_{R,zx}^{y,e} \tilde{\alpha}_{R,zx}^{y,h} + Q \ ,$$

$$\epsilon_2 = \tilde{\alpha}_{R,zy}^{x,e} \tilde{\alpha}_{R,zy}^{x,h} + \tilde{\alpha}_{R,zx}^{y,e} \tilde{\alpha}_{R,zx}^{y,h} - Q \ ,$$

$$\epsilon_3 = -\tilde{\alpha}_{R,zy}^{x,e} \tilde{\alpha}_{R,zy}^{x,h} - \tilde{\alpha}_{R,zx}^{y,e} \tilde{\alpha}_{R,zx}^{y,h} + Q \ ,$$

$$\epsilon_4 = -\tilde{\alpha}_{R,zy}^{x,e} \tilde{\alpha}_{R,zy}^{x,h} - \tilde{\alpha}_{R,zx}^{y,e} \tilde{\alpha}_{R,zx}^{y,h} - Q \ ,$$

(S26)

where $Q = \sqrt{[(\tilde{\alpha}_{R,xz}^{y,e})^2 + (\tilde{\alpha}_{R,yz}^{x,e})^2][(\tilde{\alpha}_{R,xz}^{y,h})^2 + (\tilde{\alpha}_{R,yz}^{x,h})^2]}$. The corresponding eigenstates can be written up to a normalization constant as:

$$|\psi_1\rangle = \frac{i(\tilde{\alpha}_{R,xz}^{y,e} \tilde{\alpha}_{R,xz}^{y,h} + \tilde{\alpha}_{R,yz}^{x,e} \tilde{\alpha}_{R,yz}^{x,h} + Q)}{-\tilde{\alpha}_{R,yz}^{x,e} \tilde{\alpha}_{R,yz}^{x,h} + \tilde{\alpha}_{R,xz}^{y,e} \tilde{\alpha}_{R,xz}^{y,h}}|d\rangle + |z\rangle \ ,$$

$$|\psi_2\rangle = \frac{i(-\tilde{\alpha}_{R,xz}^{y,e} \tilde{\alpha}_{R,xz}^{y,h} - \tilde{\alpha}_{R,yz}^{x,e} \tilde{\alpha}_{R,yz}^{x,h} + Q)}{\tilde{\alpha}_{R,yz}^{x,e} \tilde{\alpha}_{R,yz}^{x,h} - \tilde{\alpha}_{R,xz}^{y,e} \tilde{\alpha}_{R,xz}^{y,h}}|d\rangle + |z\rangle \ ,$$

$$|\psi_3\rangle = \frac{-\tilde{\alpha}_{R,xz}^{y,e} \tilde{\alpha}_{R,xz}^{y,h} + \tilde{\alpha}_{R,yz}^{x,e} \tilde{\alpha}_{R,yz}^{x,h} - Q}{\tilde{\alpha}_{R,yz}^{x,e} \tilde{\alpha}_{R,yz}^{x,h} + \tilde{\alpha}_{R,xz}^{y,e} \tilde{\alpha}_{R,xz}^{y,h}}|x\rangle + |y\rangle \ ,$$

$$|\psi_4\rangle = \frac{-\tilde{\alpha}_{R,xz}^{y,e} \tilde{\alpha}_{R,xz}^{y,h} + \tilde{\alpha}_{R,yz}^{x,e} \tilde{\alpha}_{R,yz}^{x,h} + Q}{\tilde{\alpha}_{R,yz}^{x,e} \tilde{\alpha}_{R,yz}^{x,h} + \tilde{\alpha}_{R,xz}^{y,e} \tilde{\alpha}_{R,xz}^{y,h}}|x\rangle + |y\rangle \ .$$

(S27)

In this configuration, in which the Rashba asymmetry direction $\boldsymbol{n}$ contains components along two nanocrystal symmetry axes, that is, $\boldsymbol{n}$ lies in a mirror plane, the dark exciton is activated. It is easy to show that the directions of the dipoles of the former bright states remain orthogonal to each other, despite that their dipole orientations have been changed. The only non-orthogonal pair of dipoles here correspond to $|\psi_1\rangle$ and $|\psi_2\rangle$, but they both are polarized along the same $z$ direction.

Going further, we can consider a general orientation of the Rashba asymmetry axis. In that case, we have not yet found a closed-form solution to the expressions above. Nevertheless, it is clear that, for a general asymmetry direction $\boldsymbol{n}$, the dark state mixes with each of the bright states, creating a higher-order coupling between each state via the "dark" intermediate state. This results in the orthogonality of the dipoles being weakly broken.

Numerical calculations have been performed that confirm this.

The results above can be understood in the context of group theory. If the Rashba asymmetry direction $\boldsymbol{n}$ is parallel to any one symmetry axis of the orthorhombic nanocrystal, the symmetry is reduced to $C_{2v}$. In that case, the dark state remains dipole inactive and the three bright states are split into mutually orthogonal, linearly polarized dipoles. But if the Rashba asymmetry also has a component along either of the other two axes, the symmetry is reduced further to $C_s$ for which no dark state exists. Finally, for a general Rashba asymmetry direction, with components along all three nanocrystal symmetry axes, the symmetry is reduced to $C_1$ for which all exciton states are coupled and as a result all dipole components are present for every state. These considerations are further discussed in Section S2.

## F. Rashba coefficient in inorganic perovskite nanocrystals

From the results from the previous subsection, Section S1.E, we now estimate the Rashba coefficients necessary to explain our experimental exciton fine-structure splittings, which are $\sim 1$ meV. Specifically, we can exploit Eq. (S25). For simplicity, we assume that all Rashba coefficients for the electron and hole are equal to each other, $\alpha_{R,jk}^{i,e} = \alpha_{R,jk}^{i,h} = \overline{\alpha}$ for any $i$, $j$, and $k$, and that their effective masses are equal: $m_e = m_h = M/2$. The Rashba energy $E_R = \overline{\alpha}^2 m_e/2$ can then be found as

$$4E_R \times (A_R^c - A_R^r) \approx 1 \text{ meV}. \tag{S28}$$

This gives $E_R \approx 0.92$ meV. For comparison, in organic perovskites it was found to be $13$ meV [25]. It is more appropriate, however, to compare the Rashba coefficient $\alpha_R$ rather than the Rashba energy between different materials, because the Rashba coefficient does not depend on the effective mass. Using $m_e = 0.13$ for CsPbBr$_3$ from Table S1, we obtain for the traditional definition of the Rashba coefficient $\alpha_R = \overline{\alpha}\hbar = 0.38$ eVÅ. For comparison, the measured value for InSb/InAsSb quantum wells is $\alpha_R = 0.14$ eVÅ[26]; in InGaAs/InP quantum wells, $\alpha_R = 0.065$ eVÅ[27], and in InAs quantum wells, $\alpha_R = 0.22$ eVÅ[28]. In organic–inorganic hybrid perovskites $\alpha_R$ is much larger due to their ferroelectricity: $\alpha_R = 7\pm1$ eVÅ in ortho-CH$_3$NH$_3$PbBr$_3$; while $\alpha_R = 11\pm4$ eVÅ in cubic-CH$_3$NH$_3$PbBr$_3$ [29]. The value of $\alpha_R$ we estimate from the experimental data is quite reasonable in comparison with other semiconductors. This leads us to conclude that the Rashba effect indeed is responsible for the observed exciton fine structure in the CsPbX$_3$ perovskite nanocrystals.

## S2. SYMMETRY ANALYSIS OF THE EXCITON FINE-STRUCTURE

Here we consider the point-group symmetry and irreducible representations appropriate for describing the band-edge excitons of quasi-cubic perovskite nanocrystals. Table S3 below shows how degeneracies and selection rules are modified as we descend in symmetry from cubic ($O_h$) to tetragonal ($D_{4h}$) or orthorhombic ($D_{2h}$) due to lattice or shape distortions. Note that in the table, the irreducible representation labels are given for the nanocrystal point group rather than the bulk space group. For each of these "parent" point groups, we also show the symmetry breaking effect of a Rashba asymmetry for different orientations $\boldsymbol{n}$ of the asymmetry axis. The groupings for each parent group ($O_h$, $D_{4h}$, and $D_{2h}$) are separated by double vertical lines in Table S3. Optical selection rules for exciton transitions are shown in Table S3 by writing the allowed transition-dipole components for each exciton irreducible representation as $x, y, z$ for linear polarized dipoles or $\sigma_{\pm}$ for circular polarization. In constructing the table, we utilized the irreducible representation labels, and the character and multiplication tables of KDWS [3].

The results summarized in the table show that the cubic perovksite band-edge exciton fine structure consists of a threefold degenerate (triplet) bright state split from a singlet dark state. As the nanocrystal symmetry is reduced by unit-cell or shape distortions the bright triplet is expected to split. For the tetragonal phase $D_{4h}$, the triplet splits into a singlet linearly polarized along the axis of symmetry and a doublet circularly polarized perpendicular to the symmetry axis. An orthorhombic distortion of $D_{2h}$ symmetry will split the bright triplet into three non-degenerate states each linearly polarized along the orthorhombic symmetry axes as follows: $\Gamma_4^-(\sigma_+, \sigma_-, z) \rightarrow \Gamma_2^-(y) + \Gamma_3^-(z) + \Gamma_4^-(x)$.

The addition of a Rashba asymmetry further breaks the symmetry of the nanocrystal beyond the shape or lattice distortions just discussed. A Rashba asymmetry directed along the $z$-axis breaks the symmetry of cubic ($O_h$) and tetragonal ($D_{4h}$) nanocrystals to $C_{4v}$, characterized by a dark singlet, a bright doublet and a linearly polarized singlet. The effect of a Rashba asymmetry along the $z$ axis of an orthorhombic nanocrystal takes the symmetry from $D_{2h}$ to $C_{2v}$, maintaining the dark state and three linearly polarized, orthogonal bright

| $O_h$ | $C_{4v}$ ($O_h$ +$n_z\hat{z}$) | $D_{4h}$ | $C_{4v}^z$ ($D_{4h}$ +$n_z\hat{z}$) | $C_{2v}^x$ ($D_{4h}$ +$n_x\hat{x}$) | $C_s^{\sigma_{xz}}$ ($C_{4v}$ +$n_x\hat{x}$) | $D_{2h}$ | $C_{2v}$ ($D_{2h}$ +$n_z\hat{z}$) | $C_s^{\sigma_{xz}}$ ($C_{2v}^z$ +$n_x\hat{x}$) | $C_1$ ($C_s^{\sigma_{xz}}$ +$n_y\hat{y}$) |
|---|---|---|---|---|---|---|---|---|---|
| $\Gamma_1^-$ (d) | $\Gamma_2$ (d) | $\Gamma_1^-$ (d) | $\Gamma_2$ (d) | $\Gamma_3$ (d) | $\Gamma_2$ (y) | $\Gamma_1^-$ (d) | $\Gamma_3$(d) | $\Gamma_2$ (y) | $\Gamma_1$ (x,y,z) |
| | | | | $\Gamma_1$(x) | $\Gamma_1$ (x, z) | $\Gamma_4^-$ (x) | $\Gamma_2$ (x) | $\Gamma_1$ (x, z) | $\Gamma_1$ (x,y,z) |
| $\Gamma_4^-$ ($z, \sigma_\pm$) | $\Gamma_5 (\sigma_\pm)$ | $\Gamma_5^- (\sigma_\pm)$ | $\Gamma_5 (\sigma_\pm)$ | $\Gamma_2$ (y) | $\Gamma_2$ (y) | $\Gamma_2^-$ (y) | $\Gamma_4$ (y) | $\Gamma_2$ (y) | $\Gamma_1$ (x,y,z) |
| | $\Gamma_1$ (z) | $\Gamma_2^-$ (z) | $\Gamma_1$ (z) | $\Gamma_4$ (z) | $\Gamma_1$ (z, x) | $\Gamma_3^-$ (z) | $\Gamma_1$ (z) | $\Gamma_1$ (z, x) | $\Gamma_1$ (x,y,z) |

excitons. However, if the Rashba asymmetry is directed off the principle axis, the symmetry reduces to $C_s$ in the case that the asymmetry is oriented within a mirror plane of the nanocrystal, and to $C_1$ otherwise. In both cases, the dark exciton state is mixed with the bright excitons. In the case of symmetry $C_s$, the Rashba asymmetry further mixes two of the bright excitons; in the lowest symmetry case, all bright excitons are mixed and the orthogonality of the dipoles is broken, at least in principle. Calculations show, however, that this mixing is a second order effect and the resulting non-orthogonality of the dipoles is expected to be weak.

Note that in Table S3, where relevant, the z-axis is taken as the principle axis unless otherwise specified as a superscript on the group symbol. The $x, y$ axes are then the axes associated with other symmetry elements such as $C_2$ rotations where they exist. When a Rashba asymmetry is oriented along a particular direction that creates a mirror plane, this mirror plane is given as a superscript on the point-group symbol. For example, $C_s^{\sigma_{xz}}$ denotes the $C_s$ point group defined with a mirror plane $\sigma_{xz}$ containing the $z$ and $x$ axes.

## S3. CALCULATION OF EXCITON AND TRION RADIATIVE LIFETIMES

### A. Radiative lifetime of excitons in cube-shaped nanocrystals

The probability of optical excitation of or recombination from any exciton state $|\Psi^{ex}\rangle$ is proportional to the square of the matrix element of the operator $\boldsymbol{e}\hat{\boldsymbol{p}}$ between that state and the vacuum state, where $\boldsymbol{e}$ is the polarization vector of the emitted or absorbed light, and $\hat{\boldsymbol{p}}$ is the momentum operator. In cube-shaped nanocrystals, the calculation of these matrix elements is complicated by the fact that the electric field of a photon inside the nanocrystal not only changes its value from outside due to dielectric screening, as in spherical nanocrystals, but it also becomes inhomogeneous.

As we discussed in Section S1.E, the triplet state in perovskite nanocrystals with orthorhombic symmetry is always split into three orthogonal dipoles with the same oscillator transition strength. Therefore, let us consider the optical transition to the triplet exciton state with $J_z = 0$, which has a linear $\boldsymbol{z}$ dipole. The square of this matrix element can be written:

$$|\langle 0|[\boldsymbol{e}^z(\boldsymbol{r})/E_\infty^z] \cdot \hat{\boldsymbol{p}}|\Psi_{1,0}^{ex}\rangle|^2 = \frac{2}{3}P^2 I_\parallel^2 \;, \tag{S29}$$

where $I_\parallel = \int d^3r\, e_z^z(\boldsymbol{r}) v(\boldsymbol{r}, \boldsymbol{r})/E_\infty^z$, and $e_z^z(\boldsymbol{r})$ is the $z$ component of the electric field of the light in the nanocrystal created by a photon with an electric field $E_\infty^z$, which is linearly polarized along $z$ and defined at infinite distance from the nanocrystal. The inhomogeneous distributions of the electric field in the cube-shaped nanocrystals created by the external homogeneous electric field is calculated in the next Section S3.B for various ratios of dielectric constants and are shown in Fig. 2e of the main text. Due to the even parity of the electron and hole envelope functions participating in the band-edge optical transitions $I_\perp = \int d^3r\, e_x^z(\boldsymbol{r}) v(\boldsymbol{r}, \boldsymbol{r})/E_\infty^z = \int d^3r\, e_y^z(\boldsymbol{r}) v(\boldsymbol{r}, \boldsymbol{r})/E_\infty^z = 0$. This analysis shows that despite the inhomogeneous distribution of the photon electric field in the cube-shaped nanocrystals the linearly polarized dipoles of each sublevel of the triplet only emit or absorb photons having a nonzero electric-field projection on the respective dipole orientation. In short, the linearly polarized dipoles emit linearly polarized light.

Substituting the matrix elements from Eq. (S29) into the expression for the radiative decay rate from ref. 30 we find the radiative lifetime of the triplet exciton $\tau_{ex}$:

$$\frac{1}{\tau_{ex}} = \frac{4e^2\omega n}{3\hbar m_0^2 c^3}|\langle 0|[\boldsymbol{e}(\boldsymbol{r})/E_\infty^z]\hat{\boldsymbol{p}}|\Psi_{1,0}^{ex}\rangle|^2 = \frac{4\omega n E_p}{9 \cdot 137 m_0 c^2} I_\parallel^2 \;, \tag{S30}$$

where $\omega$ is the transition (angular) frequency, $n$ is the refractive index of the surrounding media, $m_0$ is the free-electron mass, $c$ is the speed of light in vacuum, and $E_p = 2P^2/m_0$ is the Kane energy. The calculated radiative lifetimes in perovskite nanocrystals can be directly compared with experimental results at low temperature because the experimental data are not obscured by contributions of a low-energy dark exciton. The largest uncertainty in the radiative lifetime defined by Eq. (S30) is connected to the uncertainty in the high-frequency dielectric constant, $\epsilon_{in}$, for the perovskites, which together with $\epsilon_{out} = n^2$ determines the depolarization of the photon electric field in the nanocrystals.

Using Eq. (S30) we have calculated the radiative lifetimes in CsPbX$_3$ (X=I, Br, Cl) nanocrystals and their optical-transition energies. These are plotted in Fig. 2b in the main text. In these calculations, we used the energy-band parameters and dielectric constants from Table S1 and a refractive index of $n = 1.6$ for the surrounding medium, which yields $\epsilon_{out} = n^2$. Calculations were conducted for the three nanocrystal size regimes: (i) strong spatial confinement when the nanocrystal size $L$ is smaller than the exciton Bohr radius $a_B$, (ii) weak spatial confinement when $L \gg a_B$, and (iii) intermediate confinement when $L \sim a_B$. The discussion of these three cases is presented further in Section S3.C.

### B.  Calculation of the interior electric field in cube-shaped nanocrystals

We consider the inhomogeneous electric field inside a cube-shaped nanocrystal, modeled as a dielectric cube. The field is induced by an arbitrarily oriented electric field that is homogeneous at a large distance from the cube. Such a field can always be decomposed into three components created independently by the three projections of this remote electric field along the cube axes. In Fig. 2e in the main text, we show the distribution of the normalized $z$ component of the electric-field magnitude, $E_z^z/E_\infty^z$, in the cross-section passing through the middle of the cube, created by a homogeneous external electric field, $\boldsymbol{E}_\infty^z \parallel \boldsymbol{z}$, calculated for several ratios of internal $\epsilon_{in}$ to external $\epsilon_{out}$ dielectric constants. We describe here the approach followed to compute it.

The generalized Gauss's law states that the total electric flux through any closed surface in space of any shape drawn in an electric field is proportional to the total electric charge enclosed by the surface. The differential expression of this law, obtained via the divergence theorem, represents the local conservation of charge in the form of the well known partial differential equation (PDE)

$$\nabla \cdot (\epsilon \boldsymbol{E}) = \rho_v, \tag{S31}$$

where $\epsilon$ and $\rho_v$ represent the electric permittivity (dielectric constant) of the medium and the electric charge density, respectively. In the context of electrostatics, the electric field is computed as the negative gradient of the electric potential scalar field $\phi$, which, together with the charge conservation Eq. (S31), yields

$$\nabla \cdot (\epsilon \nabla \phi) = -\rho_v, \tag{S32}$$

which is Poisson's equation for the electric potential. This PDE is discretized and solved numerically using the finite element method (FEM). It is applied for a computational domain that involves a rectangular electrode capacitor configuration where the upper electrode is set to $\phi = 1\ V$ and the bottom one is grounded at $\phi = 0\ V$.

The proper distance between the capacitor plates in relation to the size of the embedded dielectric cube was determined by performing successive FEM analyses at various distances such that the far-field difference among all solutions remained less that 1% for various values of the dielectric constants. The distribution of both the electric potential and the electric-field magnitude along the $z$-axis line that coincides with the intersection of the $zx$ and $zy$ planes as it extends between the two electrodes of the capacitor assembly is shown in Extended Data Fig. 7 for the final selected configuration.

Contour plots of the normalized electric-field magnitude, $E_z^z/E_\infty^z$, as a function of the ratio $\epsilon_{in}/\epsilon_{out}$ are displayed in Extended Data Fig. 8. This figure shows that as the $\epsilon_{in}/\epsilon_{out}$ ratio increases, the overall magnitude of $E_z^z/E_\infty^z$ decreases. It should also be noted that the perturbations of the contours near the corners are artifacts of the interpolation resolution utilized by the software employed to construct them.

A contour plot of the normalized electric-field component $E_z^z/E_\infty^z$ on the $xz$ mid-plane for $\epsilon_{in}/\epsilon_{out}$ of 6 is shown in Fig. 2e in the main text. By symmetry this is also valid for the yz mid-plane. Contours for the normalized electric-field component $E_x^z/E_\infty^z$ for the $xz$ mid-plane are shown in Extended Data Fig. 9 for the case of $\epsilon_{in}/\epsilon_{out} = 9$. Note that the $E_y^z/E_\infty^z$ distribution on the $yz$ mid-plane is identical to Extended Data Fig. 9.

### C. The exciton-photon coupling strength in cube-shaped nanocrystals

The coupling strength between the exciton dipole and the inhomogeneous electric field of a photon in a cube-shaped nanocrystal is controlled by the square of the integral $I_\parallel$ [see Eq.(S29)]:

$$I_\parallel = \int d^3 r\, e z_z^z(\boldsymbol{r}) v(\boldsymbol{r}, \boldsymbol{r}) / E_\infty^z\ . \tag{S33}$$

In the strong-confinement regime, when the exciton Bohr radius $a_B$ is larger than the nanocrystal size $L$, then $v(\boldsymbol{r}, \boldsymbol{r}) = |\Phi_{\text{gr}}^c(\boldsymbol{r})|^2$. The ground-state wavefunctions of electrons or holes can be written as $\Phi_{\text{gr}}^c(x, y, z) = (2/L)^{3/2} \cos(\pi x/L) \cos(\pi y/L) \cos(\pi z/L)$, where $L$ is the cube edge length. Introducing dimensionless variables $\tilde{x} = x/L$, $\tilde{y} = y/L$, and $\tilde{z} = z/L$, we can rewrite $I_{\parallel}$ in dimensionless form:

$$I_{\parallel}^{strong} = 8 \int_{-0.5}^{0.5} d\tilde{x} \int_{-0.5}^{0.5} d\tilde{y} \int_{-0.5}^{0.5} d\tilde{z} [e_z^z(\tilde{x}, \tilde{y}, \tilde{z})/E_{\infty}^z] \cos^2(\pi\tilde{x}) \cos^2(\pi\tilde{y}) \cos^2(\pi\tilde{z}) . \quad (S34)$$

This expression was used in Eq. (S30) to calculate the radiative lifetimes of CsPbX$_3$ (X=I, Br, and Cl) nanocrystals with $L = 6$ and $7$ nm, $L = 5$ and $6$ nm, and $L = 4$ and $5$ nm, respectively. The energy of the optical transition in this limit is well described by $\hbar\omega = E_g + (3\hbar^2\pi^2/2\mu L^2) - 3.05e^2/\epsilon_{in}L$, where $E_g$ is the bulk energy gap of the perovskite and $\mu = (1/m_e + 1/m_h)^{-1}$ is the reduced effective mass of the exciton. We used the energy-band parameters of the perovskites from Table S1.

In the weak-confinement regime, when $L \gg a_B$, the exciton wavefunction can be written as a product of the relative motion of the electron and hole $\phi(\boldsymbol{r}_e - \boldsymbol{r}_h)$ and the exciton center-of-mass motion confined in the nanocrystal, $\Phi_{\text{gr}}^c(\boldsymbol{R})$, where $\boldsymbol{R} = (m_e\boldsymbol{r}_e + m_h\boldsymbol{r}_h)/M$ and $M = m_e + m_h$ is the exciton effective mass [9]: $v(\boldsymbol{r}_e, \boldsymbol{r}_h) = \phi(\boldsymbol{r}_e - \boldsymbol{r}_h)\Phi_{\text{gr}}^c(\boldsymbol{R})$. In the weak-confinement regime it is more convenient to directly calculate $I_{\parallel}^2$:

$$(I_{\parallel}^{weak})^2 = \frac{8}{\pi} \frac{V}{a_B^3} \left| \int_{-0.5}^{0.5} d\tilde{x} \int_{-0.5}^{0.5} d\tilde{y} \int_{-0.5}^{0.5} d\tilde{z} \frac{e_z^z(\tilde{x}, \tilde{y}, \tilde{z})}{E_{\infty}^z} \cos(\pi\tilde{x}) \cos(\pi\tilde{y}) \cos(\pi\tilde{z}) \right|^2 , \quad (S35)$$

where $V = L^3$ is the volume of the cube-shaped nanocrystal. One can see that under weak confinement, the ratio $V/a_B^3$ in $I_{\parallel}^2$ dramatically shortens the radiative decay times of the exciton in Eq.(S30) due to its giant oscillator transition strength [31]. The energy of the optical transitions in the weak-confinement regime is described as $\hbar\omega = E_g - E_B + 3\hbar^2\pi^2/2ML^2$, where $E_B = e^4\mu/(2\hbar^2\epsilon_{in}^2) = \hbar^2/(2\mu a_B^2)$ is the exciton binding energy.

Figure 2b in the main text shows the results of the calculations of the exciton radiative lifetime in perovskite nanocrystals with $L = 17$ to $25$ nm. One can see that the lifetime is strongly reduced, becoming shorter than $100$ ps in CsPbBr$_3$ and CsPbCl$_3$ nanocrystals. Further, one can see in Fig. 2b that the experimentally measured decay times are in between the lifetimes predicted for strong ($L \ll a_B$) and weak ($L \gg a_B$) confinement. This is because the correlation of the electron and hole motion in nanocrystals, which shortens the radiative decay time, can already be seen in nanocrystals with intermediate size $L \geq a_B$.

To demonstrate this, we studied the energy of the confined excitons in nanocrystals with $L \geq a_B$ using a one-parameter ansatz function:

$$v(\boldsymbol{r}_e, \boldsymbol{r}_h) = C e^{-\beta|\mathbf{r}_e - \mathbf{r}_h|} \Phi_{\text{gr}}^c(\boldsymbol{r}_e) \Phi_{\text{gr}}^c(\boldsymbol{r}_h) \ , \tag{S36}$$

where $\beta$ is a variational parameter and $C$ is a normalization constant determined by the condition $\int d^3 r_e d^3 r_h v^2(\boldsymbol{r}_e, \boldsymbol{r}_h) = 1$. Using the results of these calculations (see the next Section, Section S3.D) we show in Fig. 2b in the main text the exciton radiative lifetimes of CsPbX$_3$ (X=I, Br, and Cl) nanocrystals with $L = 6$ to 16 nm, $L = 5$ to 16 nm, and $L = 4$ to 16 nm, respectively.

### D. Variational calculation for the intermediate-confinement regime

In the variational approach we calculate the expectation value of the two-particle Hamiltonian in a cube with edge length $L$ and minimize this with respect to the variational parameter $\beta$. The Hamiltonian is

$$\hat{H} = -\frac{\hbar^2}{2m_e}\nabla_e^2 - \frac{\hbar^2}{2m_h}\nabla_h^2 - \frac{e^2}{\epsilon_{in}|\mathbf{r}_e - \mathbf{r}_h|} \ . \tag{S37}$$

Introducing the dimensionless variables $\tilde{\boldsymbol{r}}_e = \boldsymbol{r}_e/L$, $\tilde{\boldsymbol{r}}_h = \boldsymbol{r}_h/L$, and the dimensionless parameter $b = \beta L$, the expectation value $\langle v|\hat{H}|v\rangle$ reduces to the calculation of three dimensionless integrals. The first integral describes the average kinetic energy:

$$I_1(b) = \int_{-1/2}^{1/2} d^3 \tilde{r}_1 \int_{-1/2}^{1/2} d^3 \tilde{r}_2 \ e^{-b|\tilde{\mathbf{r}}_1 - \mathbf{r}_2|} \tilde{\Phi}_{\text{gr}}^c(\tilde{\mathbf{r}}_1) \tilde{\Phi}_{\text{gr}}^c(\tilde{\mathbf{r}}_2) \nabla_{\tilde{r}_1}^2 e^{-b|\tilde{\mathbf{r}}_1 - \tilde{\mathbf{r}}_2|} \tilde{\Phi}_{\text{gr}}^c(\tilde{\mathbf{r}}_1) \tilde{\Phi}_{\text{gr}}^c(\tilde{\mathbf{r}}_2) \ , \tag{S38}$$

where $\tilde{\Phi}_{\text{gr}}^c(\tilde{x}, \tilde{y}, \tilde{z}) = \cos(\pi\tilde{x})\cos(\pi\tilde{y})\cos(\pi\tilde{z})$. The second integral describes the average Coulomb interaction:

$$I_2(b) = \int_{-1/2}^{1/2} d^3 \tilde{r}_1 \int_{-1/2}^{1/2} d^3 \tilde{r}_2 \ \frac{1}{|\tilde{\mathbf{r}}_1 - \tilde{\mathbf{r}}_2|} \left( e^{-b|\tilde{\mathbf{r}}_1 - \tilde{\mathbf{r}}_2|} \tilde{\Phi}_{\text{gr}}^c(\tilde{\mathbf{r}}_1) \tilde{\Phi}_{\text{gr}}^c(\tilde{\mathbf{r}}_2) \right)^2 \ . \tag{S39}$$

Finally, the third integral determines the normalization constant $C$:

$$I_3(b) = \int_{-1/2}^{1/2} d^3 \tilde{r}_1 \int_{-1/2}^{1/2} d^3 \tilde{r}_2 \ \left( e^{-b|\tilde{\mathbf{r}}_1 - \tilde{\mathbf{r}}_2|} \tilde{\Phi}_{\text{gr}}^c(\tilde{\mathbf{r}}_1) \tilde{\Phi}_{\text{gr}}^c(\tilde{\mathbf{r}}_2) \right)^2 \ . \tag{S40}$$

The normalization constant is connected with this integral as $C = L^{-3}/\sqrt{I_3(b)}$.

Using the integrals defined in Eqs. (S38), (S39), and (S40), we can rewrite the expectation value of the Hamiltonian as,

$$\langle\Psi(b)|H|\Psi(b)\rangle = -\frac{\hbar^2}{m_e L^2}\left[\frac{1}{2}\left(1+\frac{m_e}{m_h}\right)\frac{I_1(b)}{I_3(b)}+\frac{L}{a_e}\frac{I_2(b)}{I_3(b)}\right]\ ,\qquad(S41)$$

where $a_e$ is the electron Bohr radius: $a_e = \epsilon_{in}\hbar^2/(m_e e^2)$. We find the dependence of all three integrals on $b$ using Monte-Carlo integration and determine the value of $b$ that minimizes the energy for a range of the ratios $L/a_e$. The results of these calculations are shown in Extended Data Fig. 10a.

Now we can calculate $I_\parallel$, which is defined as

$$I_\parallel = \int d^3 r[e_z^z(\mathbf{r})/E_\infty^z]v(\mathbf{r},\mathbf{r})\ .\qquad(S42)$$

Using the ansatz function definition in Eq.(S36) and the relation of the normalization constant $C$ with $I_3(b)$, we can directly connect $I_\parallel^{inter}$ in the intermediate-confinement regime with the corresponding result in the strong-confinement limit, $I_\parallel^{strong}$, defined in Eq.(S34):

$$I_\parallel^{\text{inter}}\left(\epsilon_{\text{in}}/\epsilon_{\text{out}},b\right) = \frac{1}{8\sqrt{I_3(b)}}I_\parallel^{\text{str}}\left(\epsilon_{\text{in}}/\epsilon_{\text{out}}\right)\ .\qquad(S43)$$

The radiative lifetime is proportional to $I_\parallel^2$. To describe the dependence of the radiative lifetime on the nanocrystal size, $L$, we plot the dependence $|I_\parallel^{\text{inter}}/I_\parallel^{\text{strong}}|^2$ as a function $L/a_e$ in Extended Data Fig. 10b.

## E. Radiative lifetime in spherical nanocrystals

It is interesting to compare the radiative lifetimes obtained for spherical and cube-shaped nanocrystals. In spherical nanocrystals, the electric field of the photon is homogeneous and the ratio $\mathbf{e}(\mathbf{r}_e)/E_\infty$ at each point $\mathbf{r}$ of the nanocrystal is equal to the depolarization factor $\mathcal{D}=3\epsilon_{\text{out}}/(2\epsilon_{\text{out}}+\epsilon_{\text{in}})$. Substituting this ratio into Eq. (S30) we find the radiative lifetime of the triplet exciton $\tau_{ex}$ in spherical nanocrystals:

$$\frac{1}{\tau_{ex}} = \frac{4\omega n E_p}{9\cdot 137 m_0 c^2}\mathcal{D}^2 K,\qquad(S44)$$

where $K = |\int d^3 r v(\mathbf{r},\mathbf{r})|^2$ is the overlap integral squared.

The radiative lifetime defined in Eq. (S44) depends strongly on the nanocrystal radius $a$ via the size dependence of the overlap integral $K$. In small nanocrystals that are in the strong-confinement regime ( $a < a_B$ ), the photoluminescence is determined by the optical transitions between the ground quantum confinement levels of the electrons and holes [9]. In this case, the exciton wavefunction $v(\boldsymbol{r}, \boldsymbol{r}) = \Phi_{\mathrm{gr}}^2(\boldsymbol{r})$ is the product of the two identical wavefunctions $\Phi_{\mathrm{gr}}^s(\boldsymbol{r}) = \sqrt{1/2\pi a} \sin(\pi r/a)/r$ for the electron and hole ground states resulting in $K = 1$ independent of size. The radiative lifetime in Eq.(S44) also weakly depends on nanocrystal size via the size dependence of the transition frequency, $1/\tau_{ex} \propto \omega$ .

In the weak-confinement regime, when the nanocrystal radius $a \gg a_B$ , shortening of the radiative lifetime is expected at low temperatures due to the giant oscillator transition strength of the exciton localized in the nanocrystal [31]. As discussed in Section S3.C, the exciton wavefunction can then be written as a product of the relative motion of the electron and hole $\phi(\boldsymbol{r}_e - \boldsymbol{r}_h)$ and the exciton center-of-mass motion confined in the nanocrystal. For spherical nanocrystals, the latter is $\Phi_{\mathrm{gr}}^s(\boldsymbol{R})$ yielding: $v(\boldsymbol{r}_e, \boldsymbol{r}_h) = \phi(\boldsymbol{r}_e - \boldsymbol{r}_h)\Phi_{\mathrm{gr}}^s(\boldsymbol{R})$ . Substituting this wavefunction into the overlap integral gives [9]: $K = (8/\pi^2)(a/a_B)^3$ . The resulting dramatic increase of $K$ is due to the correlation of the electron and hole motion, increasing the oscillator transition strength and shortening of the radiative lifetime. This shortening can already be observed in nanocrystals with radius $a \geq a_B$ .

The ratio of the exciton radiative lifetime in spherical and cube-shaped nanocrystals that have the same volume is equal to $I_\parallel^2/K\mathcal{D}^2$ . The result of this comparison is shown in the inset of Fig. 2e in the main text for nanocrystals in the strong- and the weak-confinement regimes. One can see that the exciton radiative lifetime in spherical nanocrystals is always shorter than in cube-shaped nanocrystals of the same volume. This is because the electric field of the photon in spherical nanocrystals is larger than the average field in cube-shaped nanocrystals.

### F. Trion radiative lifetime and polarization

The electron spin is not conserved during optical transitions in perovskite nanocrystals. As a result the trion optical transition rate is given by summing over the two possible radiative transitions the trion can undergo. The rate is the same for both positive and negative trions. Using the notation introduced in the main text, we can write for the positive trion in the strong-confinement regime:

$$\frac{1}{\tau_{\text{trion}}} = \frac{4e^2\omega n}{3\hbar m_0^2 c^3}\left(|\langle\Psi_\Uparrow^e|[e_z^z(\boldsymbol{r}_e)/E_\infty^z]\hat{p}_z|\Psi_\uparrow^h\rangle|^2 + |\langle\Psi_\Uparrow^e|e_x^x(\boldsymbol{r}_e)/E_\infty^x]\hat{p}_x||\Psi_\downarrow^h\rangle|^2\right.$$
$$\left. + |\langle\Psi_\Uparrow^e|e_y^y(\boldsymbol{r}_e)/E_\infty^y]\hat{p}_y||\Psi_\downarrow^h\rangle|^2\right) \ . \tag{S45}$$

The three matrix elements in Eq.(S45) describe transitions that are accomplished by emission of photons with three different orthogonal polarizations. All these matrix elements are equal to each other, and as a result, the trion photoluminescence is not polarized. This can be confirmed by symmetry analysis. For nanocrystals of cubic symmetry (point group $O_h$), electrons and holes have symmetry $\Gamma_6^\mp$, respectively, and are two-fold degenerate. Therefore, a positive/negative trion has symmetry $\Gamma_6^\mp$ and is also two-fold degenerate, since $\Gamma_6^+ \times \Gamma_6^- \times \Gamma_6^\pm = \Gamma_6^\mp$. Moreover, optical decay from a trion state to a single-carrier state is allowed for all polarizations: The x, y, and z components of the dipole operator all transform as $\Gamma_4^-$, and $\Gamma_6^\mp \times \Gamma_4^-$ contains $\Gamma_6^\pm$. Furthermore, the matrix elements are equal for the $x$, $y$, and $z$ components by symmetry.

For nanocrystals with orthorhombic symmetry we arrive at similar conclusions: Electrons and holes have symmetry $\Gamma_5^\mp$ respectively; the positive/negative trion has symmetry $\Gamma_5^\mp$ since $\Gamma_5^+ \times \Gamma_5^- \times \Gamma_5^\pm = \Gamma_5^\mp$. Again, dipole transitions are allowed for all polarizations since $x$, $y$, and $z$ transform as $\Gamma_4^-$, $\Gamma_2^-$, and $\Gamma_3^-$, respectively, and $\Gamma_5^\mp \times \Gamma_n^- = \Gamma_5^\pm$ for $n = 2$, 3, or 4. Given that the initial and final states involved in these transitions are twofold degenerate, it is also clear by symmetry that the trion emission has no fine structure.

Expressing the matrix elements via $I_\parallel$, we obtain for the trion lifetime

$$\frac{1}{\tau_{\text{trion}}} = \frac{2\omega n E_p}{3 \cdot 137 m_0 c^2}\left(I_\parallel^{str}\right)^2 . \tag{S46}$$

Comparison of this expression with the exciton radiative lifetime in the strong-confinement regime from Eq. (S30) shows that trion lifetime is shorter: $\tau_{\text{trion}} = (2/3)\tau_{ex}$.

In Fig. 2b in the main text we show the experimental decay times measured in perovskite nanocrystals via single-nanocrystal experiments. The photoluminescence traces shown in Fig. 2c,d demonstrate A-type blinking [32], during which the drops in photoluminescence intensity are correlated with shortening of the decay time. The intermittency of the photoluminescence intensity and the decay time is connected with switching between trion and exciton emission as a result of nanocrystal charging. As a result Fig. 2 provides direct information on the exciton radiative lifetime. The radiative decay time of the trion is 1.5 times faster than that of the exciton. The $\sim 2.5$-fold drop in photoluminescence intensity sug-gests however that for these nanocrystals, non-radiative Auger recombination significantly contributes to the decay, further shortening the trion lifetime.

Another indicator to distinguish between exciton transitions and those from trions is their polarization dependence. According to Fig. 3a-c, we observe that excitonic transitions exhibit typical dipolar emission with a high degree of linear polarization. In Extended Data Fig. 2a we plot the spectrum of a single CsPbBr$_2$Cl nanocrystal that exhibits two emission peaks at 2.5158 and 2.5175 eV and an additional trion-emission peak that is red-shifted by approximately 16 meV. For measuring the polarization, we analyzed the emitted intensity of both exciton peaks and the trion peak as function of the linear polarizer angle in front of the spectrograph, which can be seen in Extended Data Fig. 2b. Both exciton peaks, depicted in blue and red, show again typical dipolar emission along different axes of polarization, which are indicated by the blue and red straight lines. The trion peak is essentially unpolarized, in agreement with theory.

## S4. RELATIVE INTENSITIES OF PHOTOLUMINESCENCE FROM THREE ORTHOGONAL DIPOLES AND THEIR POLARIZATION PROPERTIES

The relative intensity of the photoluminescence created by three orthogonal emitting dipoles polarized along the $\boldsymbol{x}$, $\boldsymbol{y}$, and $\boldsymbol{z}$ axes and its polarization properties depend on the observation direction. The probability of emitting light for each of the dipoles is proportional to $\propto |\mathbf{e} \cdot \mathbf{x}|^2$, $\propto |\mathbf{e} \cdot \mathbf{y}|^2$, and $\propto |\mathbf{e} \cdot \mathbf{z}|^2$, where the polarization unit vector $\mathbf{e}$ is perpendicular to the light-propagation direction. To calculate these dependences we denote the light-propagation direction by the unit vector $\mathbf{u}$ with components given by

$$
\begin{aligned}
u_x &= \sin\theta\cos\phi \ , \\
u_y &= \sin\theta\sin\phi \ , \\
u_z &= \cos\theta \ ,
\end{aligned}
\tag{S47}
$$

where $\theta, \phi$ are the standard polar angles. We then form a light polarization unit vector $\mathbf{e}$ in the plane perpendicular to $\mathbf{u}$ by

$$
\mathbf{e} = (0, -u_z, u_y)/\sqrt{u_z^2 + u_y^2} \ .
\tag{S48}
$$

Finally, we gradually rotate the vector $\mathbf{e}$ around $\mathbf{u}$ and calculate the scalar products of the form $\propto |\mathbf{e} \cdot \mathbf{x}|^2$, $\propto |\mathbf{e} \cdot \mathbf{y}|^2$, $\propto |\mathbf{e} \cdot \mathbf{z}|^2$ at each angle. Each one of these scalar products

represents the probability that the corresponding dipole will emit light in the direction $\mathbf{u}$ with linear polarization $\mathbf{e}$.

The rotation of $\mathbf{e}$ around $\mathbf{u}$ is performed using the following transformation

$$\begin{pmatrix} e_x \\ e_y \\ e_z \end{pmatrix} \rightarrow \begin{pmatrix} Cs + u_x^2(1-Cs) & u_x u_y(1-Cs) - u_z Si & u_x u_z(1-Cs) + u_y Si \\ u_y u_x(1-Cs) + u_z Si & Cs + u_y^2(1-Cs) & u_y u_z(1-Cs) - u_x Si \\ u_z u_x(1-Cs) - u_y Si & u_z u_y(1-Cs) + u_x Si & Cs + u_z^2(1-Cs) \end{pmatrix} \begin{pmatrix} e_x \\ e_y \\ e_z \end{pmatrix},$$

(S49)

where $Cs = \cos\alpha$, $Si = \sin\alpha$, and $\alpha$ is a rotation angle. In fact, our calculations directly simulate the measurements performed by placing a linear polarizer perpendicular to a certain direction with respect to the emitting dipoles, and recording the intensity of the transmitted light as a function of the polarizer angle for each of the dipoles.

To obtain the total photoluminescence intensity emitted in the direction $\mathbf{u}$ for each of the lines, we integrate $|\mathbf{e}\cdot\mathbf{x}|^2$, $|\mathbf{e}\cdot\mathbf{y}|^2$, and $|\mathbf{e}\cdot\mathbf{z}|^2$ over all polar angles $\alpha$. In Extended Data Fig. 11 we provide several examples of the angular dependence of the emission probability in the plane perpendicular to the light-propagation direction and the relative photoluminescence intensity which can be observed from three orthogonal emitting dipoles, for four different directions. The calculations were conducted for: (i) a high temperature, $T$, that results in equal occupation of all three exciton levels (Extended Data Fig. 11a-d) and (ii) a temperature $T$ that provides thermal energy that is comparable to the fine-structure splitting, $kT = \Delta_1 = \Delta_2$ (Extended Data Fig. 11e-h).

One can see in Extended Data Fig. 11a-d that two dipoles contribute for any observation direction. The one photoluminescence line can be observed only if the upper exciton sublevels are unoccupied (compare photoluminescence spectra in panels a and b with the ones in panels e and f). The relative photoluminescence intensities of two lines whose polarizations are perpendicular to each other allows us to measure the relative population of the exciton spin sublevels and therefore the effective temperature (compare panels b and f). One can also see that at high temperature when all exciton sublevels are populated the detected photoluminescence intensity of the upper energy line can be larger than that of the lower energy line (compare panels d and h).

Extended Data Fig. 11 shows photoluminescence intensity peaks and their polarization calculated for cube-shaped nanocrystals. In perovskite nanostructures with orthorhombic symmetry the triplet exciton state is always split into three orthogonal dipoles. As a result, the polarization curves should look very similar to the curves shown in the insets of Extended Data Fig. 11. The intensity of photoluminescence emitted by each of these dipoles can be very sensitive to the nanocrystal shape, due to the different screening of the photon electric field by the different facets of nanocrystals with non-cube shapes. The fluctuation of the nanocrystal shape could also affect the radiative decay time of the nanocrystals.

Varying the observation directions we can describe the photoluminescence polarization curves measured in individual-nanocrystal experiments (compare insets in Fig. 3a-c and Fig. 3d-f in the main text) and determine the observation direction. To describe these photoluminescence polarization curves we assumed a Boltzmann occupation of the exciton spin sublevels with $T = 20$ K.

[S1] Bir, G. L. & Pikus, G. E. *Symmetry and Strain-Induced Effects in Semiconductors* (Wiley, New York, 1974).

[S2] Efros, Al. L. & Rosen, M. The electronic structure of semiconductor nanocrystals. *Annu. Rev. Mater. Sci.* **30**, 475-521 (2000).

[S3] Koster G. F., Dimmock, J. O., Wheeler R. G. & Statz, H. *Properties of the Thirty-Two Point Groups* (MIT Press, Cambridge, 1963).

[S4] Kane, E. O. in *Semiconductors and Semimetals*, edited by R. K. Willardson & A. C. Beer (Academic Press, New York, 1966), Vol. 1, pp. 75-100.

[S5] Yu, P. Y. & Cardona, M. *Fundamentals of Semiconductors* (Springer-Verlag, Berlin, 2001).

[S6] Pazhuk, I. P., Pydzirailo, N. S. & Matsko, M. G. Exciton absorption, luminescence and resonant Raman scattering of light in perovskite $CsPbCl_3$ and $CsPbBr_3$ crystals at low temperature. *Sov. Phys. Sol. State* **23**, 1263-1265 (1981).

[S7] Stoumpos, C. C., Malliakas, C. D. & Kanatzidis, M. G. Semiconducting tin and lead iodide perovskites with organic cations: Phase transitions, high mobilities, and near-infrared photoluminescent properties. *Inorg. Chem. 2013* **52**, 9019-9038.

[S8] Sendner, M., Nayak, P. K., Egger, D. A., Beck, S., Müller, C., Epding, B., Kowalsky, W., Kronik, L., Snaith, H. J., Pucci, A. & Lovrincic, R. Optical phonons in methylammonium lead halide perovskites and implications for charge transport. *Mater. Horiz.* **3**, 613-620 (2016).

[S9] Efros, Al. L. & Efros, A. L. Interband absorption of light in semiconductor sphere. *Sov. Phys.*

*Semicond.* **16**, 772-775 (1982).

[S10] Brus, L. E. Electron-electron and electron-hole interactions in small semiconductor crystal-lites: The size dependence of the lowest excited electronic state. *J.Chem. Phys.* **80**, 4403-4409 (1984).

[S11] Nirmal, M., Norris, D. J., Kuno, M., Bawendi, M. G., Efros, Al. L. & Rosen, M. Observation of the 'dark exciton' in CdSe quantum dots. *Phys. Rev. Lett.* **75**, 3728-3731 (1995).

[S12] Tanaka, K., Takahashi, T., Ban, T., Kondo, T., Uchida, K. & Miura, N. Comparative study on the excitons in lead-halide-based perovskite-type crystals $CH_3NH_3PbBr_3$ $CH_3NH_3PbI_3$. *Solid State Commun.* **127**, 619-623 (2003).

[S13] Even, J. Pedestrian guide to symmetry properties of the reference cubic structure of 3d all-inorganic and hybrid perovskites. *J. Phys. Chem. Lett.* **6**, 2238-2242 (2015).

[S14] Kataoka, T., Kondo, T., Ito, R., Sasaki, S., Uchida, K. & Miura, N. Magneto-optical study on excitonic spectra in $(C_6H_{13}NH_3)_2PbI_4$. *Phys. Rev. B* **47**, 2010-2018 (1993).

[S15] Tanaka, K., Takahashi, T., Kondo, T., Umeda, K., Ema, K., Umebayashi, T., Asai, K., Uchida, K. & Miura, N. Electronic and excitonic structures of inorganic–organic perovskite-type quantum-well crystal $(C_4H_9NH_3)_2PbBr_4$. *Jpn. J. Appl. Phys.* **44**, 5923-5932 (2005).

[S16] Fu, M., Tamarat, P., Huang, H., Even, J., Rogach, A. L. & Lounis, B. Neutral and charged exciton fine structure in single lead halide perovskite nanocrystals revealed by magneto-optical spectroscopy. *Nano Lett.* **17**, 2895-2901 (2017).

[S17] Elliott, R. J. Symmetry of excitons in $Cu_2O$. *Phys. Rev.* **124**, 340-345 (1961).

[S18] Pikus, G. E. & Bir, G. L. Exchange interaction in excitons in semiconductors. *Sov. Phys. JETP* **33**, 108-114 (1971).

[S19] Bassani, F., Forney, J. J. & Quattropani, A. Energy level structure of biexcitons and related optical transitions. *Phys. Stat. Sol. B* **65**, 591-601 (1974).

[S20] Cottingham, P. & Brutchey, R. L. On the crystal structure of colloidally prepared $CsPbBr_3$ quantum dots. *Chem. Commun.* **52**, 5246-5249 (2016).

[S21] Yaffe, O., Guo, Y., Tan, L. Z., Egger, D. A., Hull, T., Stoumpos, C. C., Zheng, F., Heinz, T. F., Kronik, L., Kanatzidis, M. G., Owen, J. S., Rappe, A. M., Pimenta, M. A. & Brus, L. E. Local polar fluctuations in lead halide perovskite crystals. *Phys. Rev. Lett.* **118**, 136001 (2017).

[S22] Bychkov, Yu. A. & Rashba, E. I. Oscillatory effects and the magnetic susceptibility of carriers

in inversion layers. *J. Phys. C: Solid State Phys.* **17** 6039-6045 (1984).

[S23] Landau, L. D. & Lifshitz, E. M. *Quantum Mechanics* (Pergamon Press, Oxford, 1960).

[S24] Winkler, R. *Spin–Orbit Coupling Effects in Two-Dimensional Electron and Hole Systems* (Springer-Verlag, Berlin-Heidelburg, 2003).

[S25] Kim, M., Im, J., Freeman, A. J., Ihm, J. & Jin, H. Switchable S = 1/2 and J = 1/2 Rashba bands in ferroelectric halide perovskites. *Proc. Natl. Acad. Sci.* **111** 6900-6904 (2014).

[S26] Gilbertson, A. M., Branford, W. R., Fearn, M., Buckle, L., Buckle, P. D., Ashley, T. & Cohen, L. F. Zero-field spin splitting and spin-dependent broadening in high-mobility InSb/In$_{1-x}$Al$_x$Sb asymmetric quantum well heterostructures. *Phys. Rev. B* **79**, 235333 (2009).

[S27] Guzenko, V. A., Schäpers, T. & Hardtdegen, H. Weak antilocalization in high mobility Ga$_x$In$_{1-x}$As/InP two-dimensional electron gases with strong spin–orbit coupling. *Phys. Rev. B* **76**, 165301 (2007).

[S28] Grundler, D. Large Rashba splitting in InAs quantum wells due to electron wave function penetration into the barrier layers. *Phys. Rev. Lett.* **84**, 6074-6077 (2000).

[S29] Niesner, D., Wilhelm, M., Levchuk, I., Osvet, A., Shrestha, S., Batentschuk, M., Brabec, C. & Fauster, T. Giant Rashba splitting in CH$_3$NH$_3$PbBr$_3$ organic–inorganic perovskite. *Phys. Rev. Lett.* **117**, 126401 (2016).

[S30] Landau, L. D. & Lifshitz, E. M. *Electrodynamics of Continuous Media* (Pergamon Press, Oxford, 1960).

[S31] Rashba, E. I. & Gurgenishvili, G. E. Edge absorption theory in semiconductors. *Sov. Phys. Solid State* **4**, 759-760 (1962).

[S32] Galland, C., Ghosh, Y., Steinbrück, A., Sykora, M., Hollingsworth, J. A., Klimov, V. I. & Htoon H. Two types of luminescence blinking revealed by spectroelectrochemistry of single quantum dots. *Nature* **479**, 203-207 (2011).